\documentclass[journal]{IEEEtran}
\IEEEoverridecommandlockouts
\usepackage{caption}
\usepackage{booktabs}
\usepackage{siunitx}
\usepackage{cite}
\usepackage{amsmath,amssymb,amsfonts}
\usepackage{graphics, framed}
\usepackage{graphicx}
\usepackage{epsfig}
\usepackage{epstopdf}
\usepackage{url}
\usepackage{multimedia}
\usepackage{paralist}
\usepackage[printonlyused]{acronym}
\usepackage{units}
\usepackage{algorithmic}
\usepackage{textcomp}
\usepackage{float}
\usepackage{tabularx}
\usepackage{balance}
\usepackage{subfigure}
\usepackage{xcolor}
\def\BibTeX{{\rm B\kern-.05em{\sc i\kern-.025em b}\kern-.08em
T\kern-.1667em\lower.7ex\hbox{E}\kern-.125emX}}
\usepackage{mathtools}
\usepackage{cuted} 
\usepackage[export]{adjustbox}

\usepackage{dblfloatfix}
\usepackage{lipsum}
\usepackage{color,soul}
\usepackage{multirow}
\usepackage[overload]{textcase}
\newcommand{\FGR}[1]{Fig.~\ref{#1}}

\newcommand{\SEC}[1]{Section~\ref{#1}}
\newcommand{\TAB}[1]{Table~\ref{#1}}
\newcommand{\EQ}[1]{(\ref{#1})}
\newcommand{\E}{\mathbb{E}}
\newcommand{\Var}{\mathrm{Var}}

\acrodef{5G}[5G]{5\textsuperscript{th}--Generation}
\acrodef{AF}[AF]{amplitude--and--forward}
\acrodef{AWGN}[AWGN]{additive white Gaussian noise}
\acrodef{BEP}[BEP]{bit error probability}
\acrodef{BPSK}[BPSK]{binary phase shift keying}
\acrodef{BW}[BW]{bandwidth}
\acrodef{CLT}[CLT]{central limit theorem}
\acrodef{CW}[CW]{continuous wave}
\acrodef{D2D}[D2D]{device--to--device}
\acrodef{dB}[dB]{decibel}
\acrodef{dBi}[dBi]{decibel isotropic}
\acrodef{dBm}[dBm]{decibel over a milliwatt}
\acrodef{Gbps}[Gbps]{gigabit per second}
\acrodef{GHz}[GHz]{gigahertz}
\acrodef{GM}[GM]{Gamma mixture}
\acrodef{LEO}[LEO]{low Earth orbit}
\acrodef{LOS}[LOS]{line of sight}
\acrodef{MGF}[MGF]{moment generating function}
\acrodef{MIMO}[MIMO]{Multiple--input multiple--output}
\acrodef{mMIMO}[mMIMO]{massive--\ac{MIMO}}
\acrodef{NLOS}[NLOS]{non--line of sight}
\acrodef{PDF}[PDF]{probability distribution function}
\acrodef{PSK}[PSK]{phase shift keying}
\acrodef{QAM}[QAM]{quadrature amplitude modulation}
\acrodef{RIS}[RIS]{reconfigurable intelligent surface}
\acrodef{SEP}[SEP]{symbol error probability}
\acrodef{BER}[BER]{Bit error rate}
\acrodef{SNR}[SNR]{signal--to--noise ratio}
\acrodef{THz}[THz]{terahertz}
\acrodef{ISL}[ISL]{inter-satellite link}
\acrodef{QoS}[QoS]{quality of service}
\ifCLASSINFOpdf
\else
\fi

\begin{document}
\title{\textcolor{black}{Reconfigurable Intelligent Surfaces Empowered THz Communication in LEO Satellite Networks}}

% \author{Author 1, Author 2, Author 3, Author 4, Author 5, Author 6}

\author{K{\"{u}}r{\c{s}}at~Tekb{\i}y{\i}k,~\IEEEmembership{Graduate Student Member,~IEEE,} G{\"{u}}ne{\c{s}}~Karabulut~Kurt,~\IEEEmembership{Senior~Member,~IEEE,} Ali~R{\i}za~Ekti,~\IEEEmembership{Senior~Member,~IEEE,} Halim~Yanikomeroglu,~\IEEEmembership{Fellow,~IEEE} 

\thanks{K. Tekb{\i}y{\i}k is with the Department of Electronics and Communications Engineering, {\.{I}}stanbul Technical University, {\.{I}}stanbul, Turkey, e-mail: tekbiyik@itu.edu.tr}
\thanks{G. Karabulut Kurt is with the Poly-Grames Research Center, Department of Electrical Engineering, Polytechnique Montr\'eal, Montr\'eal, Canada, e-mail: gunes.kurt@polymtl.ca}

\thanks{A.R. Ekti is with the Grid Communications and Security Group, Electrification and Energy Infrastructure Division, Oak Ridge National Laboratory, Oak Ridge, TN, U.S.A., e-mail: ektia@ornl.gov. This manuscript has been authored in part by UT-Battelle, LLC, under contract DE-AC05-00OR22725 with the US Department of Energy (DOE). The publisher acknowledges the US government license to provide public access under the DOE Public Access Plan (http://energy.gov/downloads/doe-public-access-plan).}

\thanks{H. Yanikomeroglu is with the Department of Systems and Computer Engineering, Carleton University, Ottawa, Canada, e-mail: halim@sce.carleton.ca} }

\IEEEoverridecommandlockouts 
\maketitle

\begin{abstract}
The revolution in the \ac{LEO} satellite networks will \textcolor{black}{bring} changes on their communication models \textcolor{black}{and a shift} from the classical bent-pipe architectures to more sophisticated networking platforms. \textcolor{black}{Thanks to} technological advancements in microelectronics and micro-systems, the \ac{THz} band \textcolor{black}{has emerged as a strong candidate for \acp{ISL} due to its promise of high data rates.} Yet, the propagation conditions of the \ac{THz} band need to be properly modeled and controlled \textcolor{black}{with} utilizing \acp{RIS} to \textcolor{black}{leverage} their full potential. In this work, we first provide an assessment of the use of the \ac{THz} band for \acp{ISL}, and quantify the impact of misalignment fading on error performance. Then, in order to compensate \textcolor{black}{for} the high path loss associated with high carrier frequencies, and to further improve the \ac{SNR}, we propose the use of \acp{RIS} mounted on neighboring satellites to enable signal propagation. Based on a mathematical analysis of the problem, we present the error rate expressions for \ac{RIS}-assisted \acp{ISL} with misalignment fading. \textcolor{black}{Also, numerical results show that RIS can leverage the error rate performance and achievable capacity of THz ISLs.}

%Numerical results demonstrate that the proposed \ac{RIS}-\textcolor{black}{empowered} \ac{THz} communication solution \textcolor{black}{presents} significant performance improvement \textcolor{black}{with} the use of \acp{RIS}.
\end{abstract}


\IEEEpeerreviewmaketitle
\acresetall

\section{Introduction}\label{sec:intro}

\textcolor{black}{Thanks to recent advances in space technology}, satellite production and deployment costs \textcolor{black}{have been} significantly reduced. As a result, \ac{LEO} satellites \textcolor{black}{have become} an attractive \textcolor{black}{option for providing} ubiquitous and low-latency communications~\cite{woellert2011cubesats}. The unprecedented growth in \ac{LEO} satellite deployments \textcolor{black}{has opened} new horizons for the wireless communication world. For example, SpaceX, which plans to \textcolor{black}{launch} thousands of satellites in \ac{LEO} and lower orbits to cover the vast majority of the \textcolor{black}{users}, aims for seamless and low latency communication between satellites~\cite{saeed2020cubesat}. To provide ubiquitous and flexible connectivity solutions, new types of satellites supporting cooperation between satellites and multi-band support are needed. Furthermore, it is envisioned that small satellites will be key drivers for space communications owing to low production and deployment costs associated with \textcolor{black}{launching} a satellite into orbit~\cite{Akyildiz2019IEEENetwork}. The satellite networks seem to include a massive number of satellites and CubeSats. For example, SpaceX is planning to deploy $1600$ satellites for Starlink's initial phase and SpaceX additionally proposes launching extra $7518$ satellites into the orbit with an altitude of $340$ km~\cite{handley2018delay}. A cooperative communication paradigm seems to be appropriate for more efficient and effective use of a network with such a large number of satellites. Therefore, the requirements for satellite communications need to be revised and redefined to suit small satellites with relatively low power transmission capacities. \textcolor{black}{As the state-of-art technology, the \ac{THz} band has emerged as a promising solution for \acp{ISL} since it can provide high-data-rate communications due to its untapped wide bandwidth~\cite{tekbiyik2020holistic, chaccour2022seven}. However, \ac{THz} waves suffer from severe path loss~\cite{schneider2012link, tekbiyik2019statistical}. Fortunately, path-loss due to molecular absorption is a non-issue for space applications of THz communications. Nevertheless, misalignment between transmitter and receiver antennas can dramatically decrease received power, as shown in~\cite{priebe2012affection, ekti2017statistical} due to their narrow beams and high directivity.}

\textcolor{black}{Furthermore,} taking the transmission power limits of \ac{LEO} satellites into account, energy-efficient communications involving THz links can be met by using software controlled surfaces~\cite{liaskos2018new}. In a recent study, transmission through \acp{RIS} was proposed as a novel communication technology with considerable potential~\cite{wu2019towards}. \textcolor{black}{The \acp{RIS} with a massive number of passive elements on a flat surface can manipulate the propagation medium by separately adjusting the phase of impinging signal by each element of RIS. The main advantage of the \acp{RIS} is the lack of active elements, which consume power, while \ac{MIMO} and relay-based communication systems employ several active elements.The passive reflectarrays and phased arrays have been already employed in satellite communication systems to indemnify the path loss through the long propagation distance. But, it should be noted that reflectarrays are not capable to adaptively adjust the characteristics of the impinging signal.  On the other hand, RISs can adaptively change beamforming direction and eliminate multipath effects~\cite{arslan2021over}.} Furthermore, the \acp{RIS} do not need complex processing or coding~\cite{basar_wireless_2019}. \textcolor{black}{In this respect, these appealing features make RISs prominent to improve the communication performance and \ac{QoS} for small satellites with low power consumption requirements.} As detailed in~\cite{basar2019transmission}, \acp{RIS} can provide energy efficiency for the same \ac{QoS} level. \textcolor{black}{RISs are expected to be a key enabler for \ac{THz} waves in small satellites of the future~\cite{ye2021non} due to the energy efficiency~\cite{tekbiyik2021energy} and low hardware complexity~\cite{khan2022ris} provided by \acp{RIS}. In~\cite{zheng2022intelligent}, an architecture for RIS-assisted LEO satellite communications with RISs deployed at satellite and ground nodes and cooperative passive beamforming are proposed. The authors jointly optimize both beamforming at satellite and ground nodes to maximize the channel gain between satellite and ground nodes. But, the misalignment fading is not considered in that study.}

In this study, we investigate \ac{RIS}-assisted \ac{THz} band \acp{ISL} in \ac{LEO} constellations, as depicted in \FGR{fig:system_model}, to meet the requirements such as low-latency, ubiquitous connectivity, high data rates, and low-complexity. Moreover, \acp{RIS} can allow flexibility in the \textcolor{black}{direction of transmission}, which is originally restricted \textcolor{black}{to} four nearby satellites in Starlink constellations~\cite{handley2019using}. \textcolor{black}{Taking the above into account, we propose a cooperation between \ac{RIS}-assisted satellites within the same and nearby orbits, and we derive error performance expressions} to assess the potential of cooperative techniques in \ac{RIS}-assisted \ac{THz} \acp{ISL}. \textcolor{black}{The expressions we derive are then validated with simulation results}. Considering the relative motion and changing distance between satellites in different orbits, the performance analyses are carried out for the upper and lower limits by setting the closest and furthest possible distances. 

\begin{figure}
    \centering
    \includegraphics[width=\linewidth, page = 2]{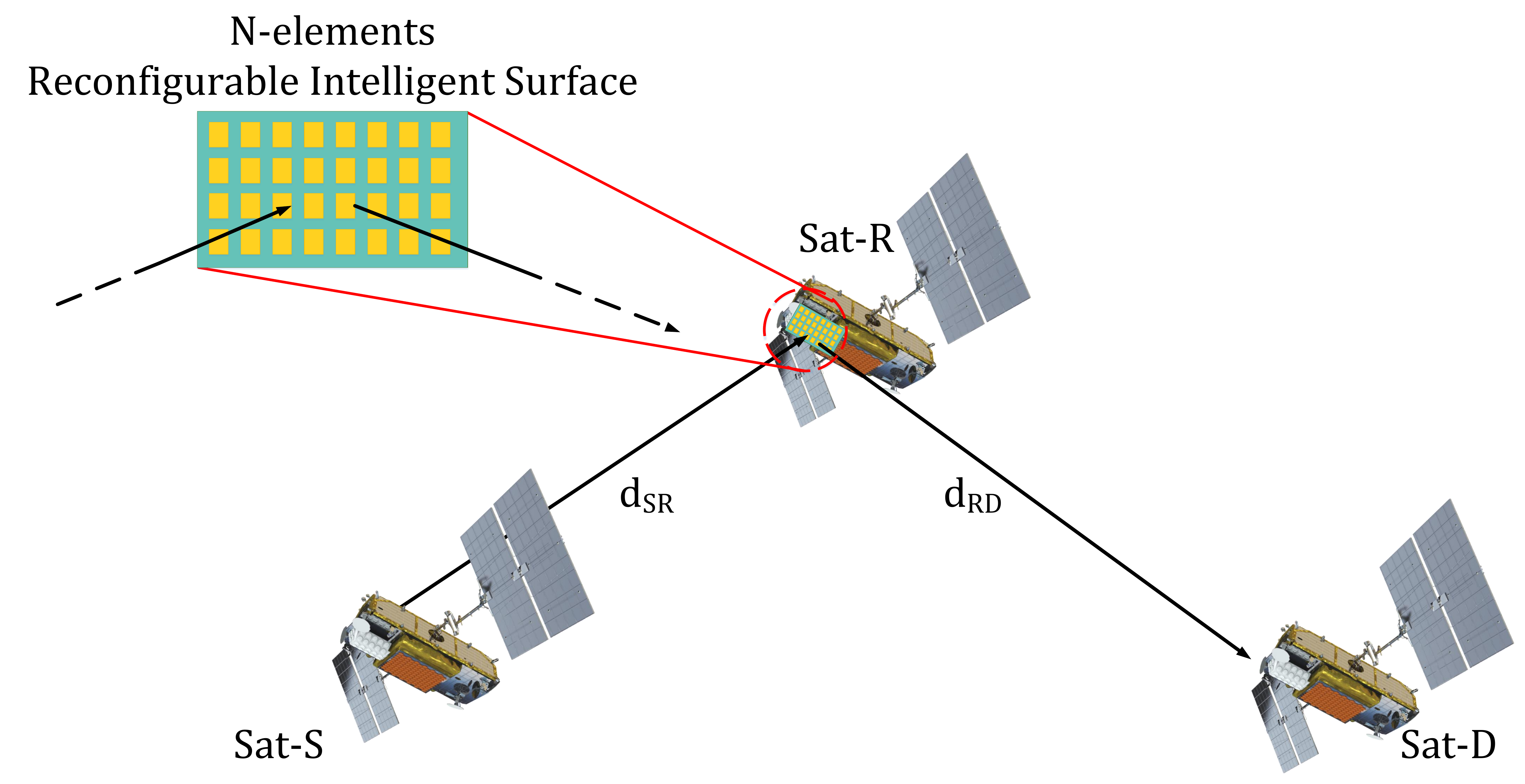}
    \caption{The system model of an RIS-empowered \ac{ISL}. The distance between the source (Sat-S) and the relay (Sat-R) is $d_{SR}$. Sat-R is equipped with an \ac{RIS} that is composed of $N$ elements. The distance between the relay and the destination (Sat-D) is $d_{RD}$.}
    \label{fig:system_model}
\end{figure}

% \subsection{Related Works}
% \subsubsection{Terahertz Wireless Communication}
% \subsubsection{Reconfigurable Intelligent Surfaces}
% \subsubsection{Low-Earth Orbit Satellites}

\subsection{Contributions}

The use of an \ac{RIS}-empowered \ac{THz} band in \ac{LEO} satellite \acp{ISL} is a promising tool to address high power consumption and low diversity order problems of a single \ac{ISL}. The contributions of this study \textcolor{black}{are summarized here:}
%the ubiquitous connection with low-latency communications problems. 

\subsubsection*{Contribution 1}
The utilization of \ac{RIS} in \ac{THz} communication systems is proposed. This enables the efficient transmission of the \ac{THz} waves, which suffer from high path loss. As shown in~\cite{yildirim2019propagation}, the effective path loss exponent can be reduced by the use of \acp{RIS} in wireless communication links. \textcolor{black}{To the best of our knowledge, this study is the first to provide an error performance analysis for RIS-assisted THz \acp{ISL} in LEO satellite networks.} 
\subsubsection*{Contribution 2}
\textcolor{black}{First}, we focus on the single \ac{RIS}-assisted \ac{THz} \acp{ISL} to enhance the power efficiency of the \ac{LEO} satellite networks, as well as to improve the achievable data rates. The authors consider the misalignment fading, which will be observed in high velocity satellite systems utilizing narrow \ac{THz} beams, in the performance analysis. To the best of \textcolor{black}{our} knowledge, this study \textcolor{black}{is the first to consider} the misalignment fading in \ac{RIS}-assisted communications.
\subsubsection*{Contribution 3}
\textcolor{black}{We address the utilization of RIS-assisted THz satellite networks in a cooperative manner in this study. The benefit of multiple satellites' cooperation is shown to enhance power efficiency for a constant bit error probability.}
\subsubsection*{Contribution 4}
Considering Starlink and Iridium as two sample systems, the performance analysis is carried out in deference to their specifications given in \TAB{tab:distances}. Thus, the performance results regarding the cases that Starlink and Iridium satellite networks employ the proposed communication framework which utilizes \acp{RIS} and \ac{THz} signaling in \acp{ISL} are revealed.

\begin{table}[!t]
\centering
\caption{The distances for Iridium and Starlink constellations.}
\begin{tabular}{lcc}
\toprule \toprule
\multicolumn{1}{l}{\textbf{Specifications}} & \multicolumn{1}{c}{\textbf{Iridium}} & \multicolumn{1}{c}{\textbf{Starlink\footnote{The values are given according to the first phase of Starlink.}}} \\  \midrule
$r_s$ (km)            &  781    &    1150    \\
$N_{\text{sat}}$      &  11     &     50     \\
$N_{\text{orbit}}$    &  6      &     32     \\ \midrule
$\theta$               &  32.73$^{\circ}$       &    7.2$^{\circ}$      \\
$\Psi$                 &  30$^{\circ}$       &    5.625$^{\circ}$     \\
$d_{\text{intra}}$ (km)   &    4034     &   945.4       \\
$d_{\text{nearest}}$ (km) &   2037.8    &   472.93       \\
$d_{\text{farthest}}$ (km) &  4162.8    &   876.57      \\ \bottomrule \bottomrule
\end{tabular}
\label{tab:distances}
\end{table}

\subsection{Outline}
The rest of this paper is organized as follows. In \SEC{sec:leo_contellations}, we introduce the associated model of the ISLs in LEO constellations in which intra-plane and inter-plane distance models are overviewed. The communication model of the RIS-assisted THz ISL links is given in \SEC{sec:system_model} along with the misalignment fading model. Also in this section, single RIS-assisted and multiple RIS-assisted communication scenarios are considered, and the associated error expressions are derived. Extensive numerical results and observations are presented in \SEC{sec:results} \textcolor{black}{including for} both single and multiple RIS aided cooperation scenarios. In \SEC{sec:open_issues}, \textcolor{black}{we present a brief discussion of related open issues. \SEC{sec:conclusion} concludes the study}.

\section{Inter-satellite Links in LEO Constellations}\label{sec:leo_contellations}

\begin{figure*}[!t]
\centering
\subfigure[]{
\label{fig:intra_distance}
\includegraphics[width=0.3\linewidth, page = 5]{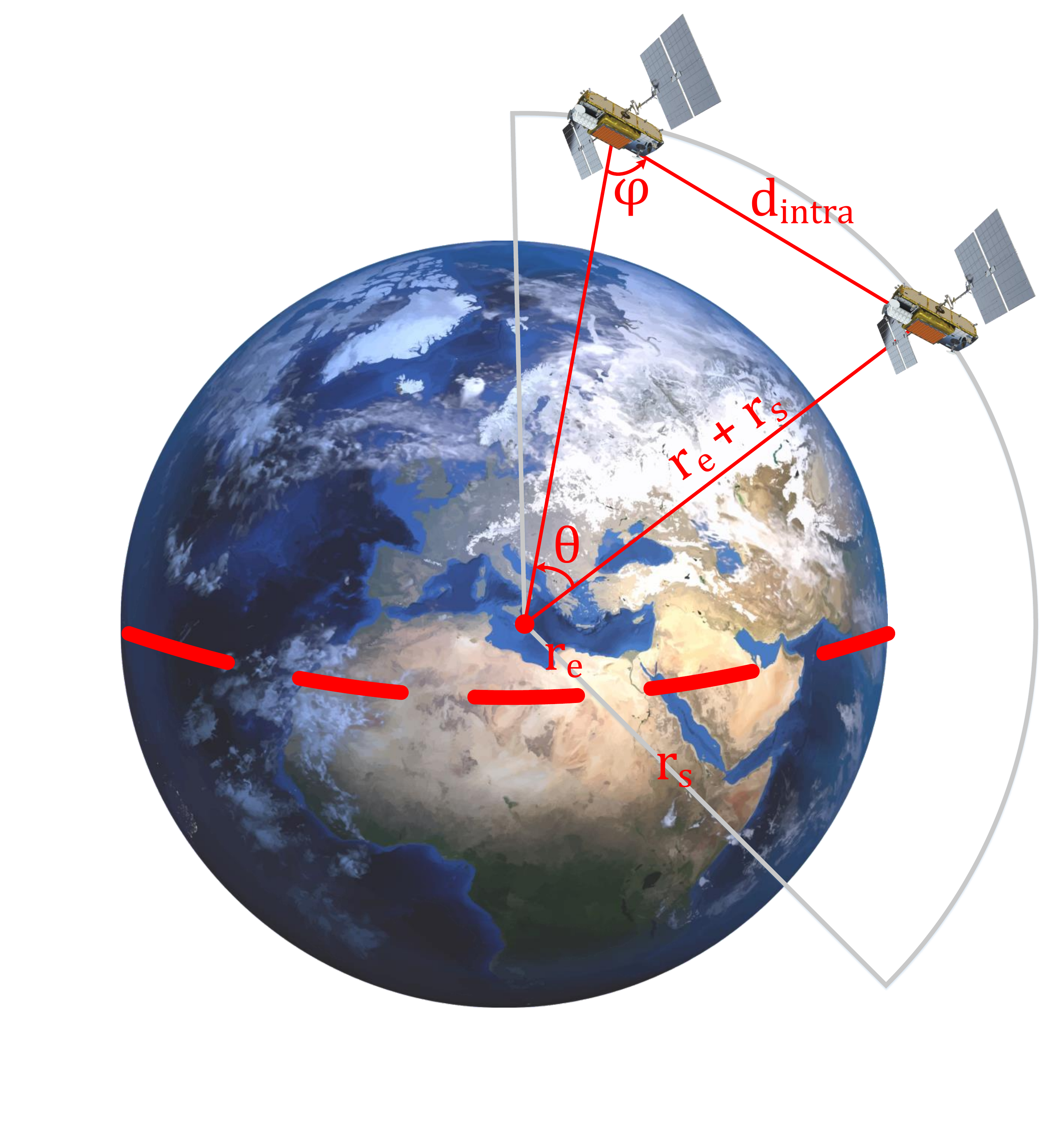}}
\subfigure[]{
\label{fig:nearest_distance}
\includegraphics[width=0.3\linewidth, page = 6]{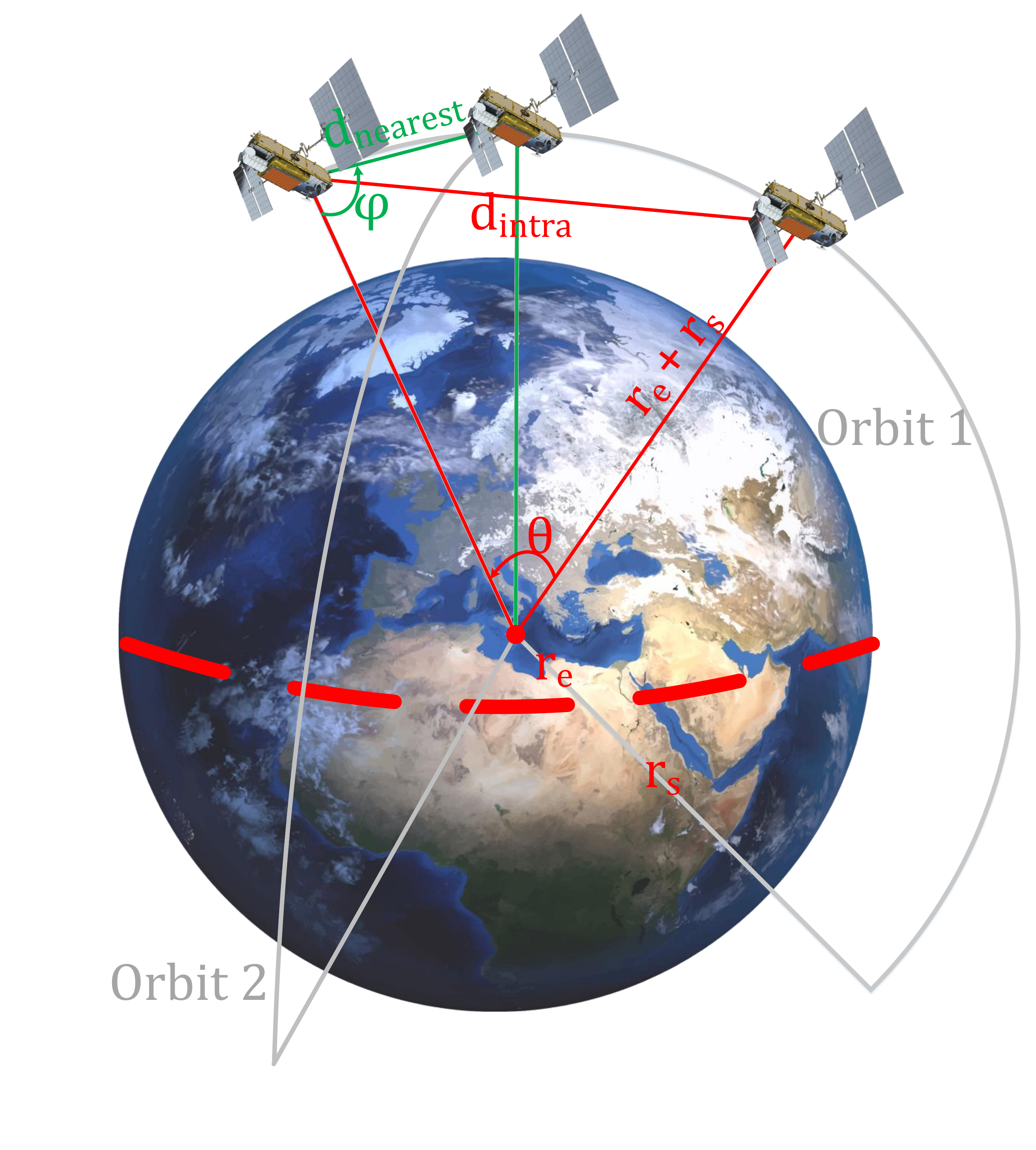}}
\subfigure[]{
\label{fig:farthest_distance}
\includegraphics[width=0.3\linewidth, page = 7]{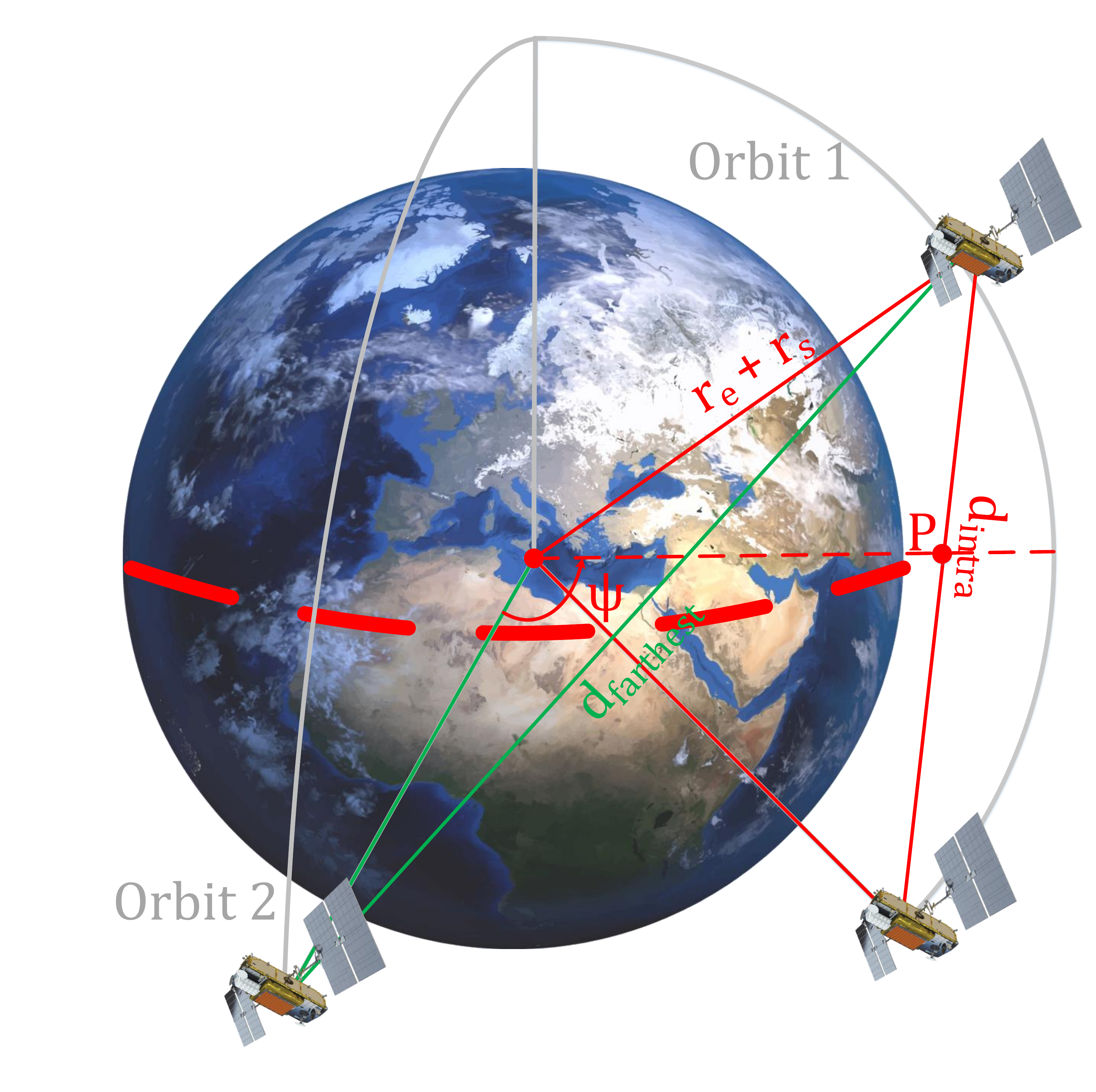}}
\caption{Satellites can be \textcolor{black}{ at various positions —and in various orbits—} relative to each other. Three of \textcolor{black}{these distances} are: (a) intra-plane distance, (b) the nearest, and (c) the farthest. While the distance between the satellites in the same orbit is constant throughout the movement, the distance between the satellites in the two neighboring orbits changes with time. Two extreme examples of the nearest and farthest are shown.}
\label{fig:satellite_distances}
\end{figure*}

\textcolor{black}{The transmission of THz waves in space results in a non-negligible loss in received power, where the distance between the transmitter and receiver is a determining parameter of this loss. This loss is proportional to the square of the distance}. \textcolor{black}{Thus, in what follows, we begin by determining the probable distance between the two satellites in both Starlink and Iridium constellations,} which are selected as sample networks for this study. 

\textcolor{black}{LEO satellites can be positioned at different distances from each other in different satellite constellations, and some of these distances can be quite significant.} \textcolor{black}{The distances between satellites} can be investigated in three cases: time-invariant distances between two neighboring satellites in the same orbit (\FGR{fig:intra_distance}), the nearest distance when one of the satellites is located just over one of the poles  (\FGR{fig:nearest_distance}), the farthest distance when one of the satellites is just over the equator (\FGR{fig:farthest_distance})~\cite{yang2019low}. \textcolor{black}{Obviously any measure of distance between two satellites} in nearby orbits lies in between $d_{nearest}$ and $d_{farthest}$. The first three cases are investigated below. In this study, we only evaluate the analysis for the maximum and minimum distances in order to determine the upper and lower limits of inter-satellite communication performance in terms of error probability. It is clear that the \textcolor{black}{distances between satellites will consist} of the values between these two distances. We \textcolor{black}{believe that a performance analysis on the basis of maximum and minimum distances} between them constantly changes due to their motion through orbits. Throughout the study, $d_{intra}$, $d_{nearest}$, and $d_{farthest}$ stand for the inter-satellite distance for an orbit, the shortest and longest distances in between two satellites in nearby orbits, respectively.

\subsection{Intra-plane \textcolor{black}{Distance}}
The satellites in the same orbit follow each other with the same distance as shown in \FGR{fig:intra_distance}. The distance \textcolor{black}{from the center of the Earth} to the satellite, with the Earth radius $r_e$, is shown by $r_e + r_s$. The angle between two neighboring satellites in the same orbit, $\theta = \frac{360^{\circ}}{N_{\text{sat}}}$, where $N_{sat}$ represents the total number of satellites in the orbit. Since the distances to Earth center are the same for each satellite, the angle between the line from center to satellite and the satellite to satellite line is calculated as 
\begin{align}
    \varphi = \frac{180^{\circ}-\theta}{2}.
\end{align}
By using the law of sines, \textcolor{black}{we obtain the following distance:}
\begin{align}
    d_{\text{intra}} = \frac{(r_e + r_s)\times\sin(\theta)}{\sin(\varphi)},
\end{align}
where $d_{\text{intra}}$ and $\varphi$ denote the distance between two neighboring satellites in the same orbit and the angle between the normal vector and the vector to the neighboring satellite as illustrated in \FGR{fig:satellite_distances}.

\subsection{Inter-plane \textcolor{black}{Distances}}
In this section, the minimum and maximum distances between satellites in nearby orbits are investigated to reveal the upper and lower limits.
\subsubsection{Nearest \textcolor{black}{Distance}}
Where two satellites in different orbits are closest to each other, one of the satellites is directly above the north or south pole. \textcolor{black}{It should be noted that, although Starlink satellites do not pass directly over either pole, the north-east orbits form virtual poles above the north and south 53rd-degree latitudes.} This scenario is illustrated in \FGR{fig:nearest_distance}. Likewise, \textcolor{black}{in the case of intra-plane distance}, the nearest distance is found by applying the law of sines as
\begin{align}
    d_{\text{nearest}} = \frac{(r_e + r_s)\times\sin(\frac{\theta}{2})}{\sin(\varphi)},
\end{align}
where $\frac{\theta}{2}$ and $\varphi = \frac{180-\theta/2}{2}$ are half of the angle between the line from center \textcolor{black}{of the Earth} to satellite and the angle between the inter-satellite line and the line through the center of the Earth, respectively.

\subsubsection{Farthest \textcolor{black}{Distance}}
When the satellite positions at just over the equator, the distance to the satellite in neighbor orbit becomes maximum as seen in \FGR{fig:farthest_distance}. The distance from the center of Earth to the point $P$ is calculated by using Pythagoras theorem as
\begin{align}
    |OP| = \sqrt{(r_e + r_s)^{2} - \left(\frac{d_{\text{intra}}}{2}\right)^{2}}.
\end{align}
Then, the distance between the satellite in Orbit 2 and the point $P$ is
\begin{align}
    d_{\text{SP}} = \sqrt{(r_e + r_s)^{2} + |OP|^{2} - 2(r_e + r_s)|OP|\cos(\Psi)}
\end{align}
where $\Psi$ is the angle between two nearby orbits. Therefore, $\Psi$ is equal to $\frac{180^{\circ}}{N_{\text{orbit}}}$ for a constellation with $N_{\text{orbit}}$ orbits.
Finally, the farthest distance between satellites is deduced as 
\begin{align}
    d_{\text{farthest}} = \sqrt{d_{\text{SP}}^2 + \left(\frac{d_{\text{intra}}}{2}\right)^{2}}.
\end{align}

As an example, the distances are shown in \TAB{tab:distances} for \textcolor{black}{the two satellite constellations considered in this study: Starlink and Iridium.} Since Starlink has more orbits and satellites per orbit, the distances are relatively short compared to the Iridium satellite network.

\section{RIS-assisted Terahertz Wireless Communication in Inter-satellite Links}\label{sec:system_model}

In this section, \ac{RIS}-assisted \ac{THz} \acp{ISL} are considered with a misalignment fading due to the sharp beams of \ac{THz} antennas and the high relative velocity of \ac{LEO} satellites. \textcolor{black}{To the best of our knowledge, this study is the first to provide an error performance analysis for RIS-assisted THz ISLs in LEO satellite networks.}

\subsection{Misalignment Fading}\label{sec:misalignment}
\begin{figure}[!t]
    \centering
    \includegraphics[width=\linewidth, page = 9]{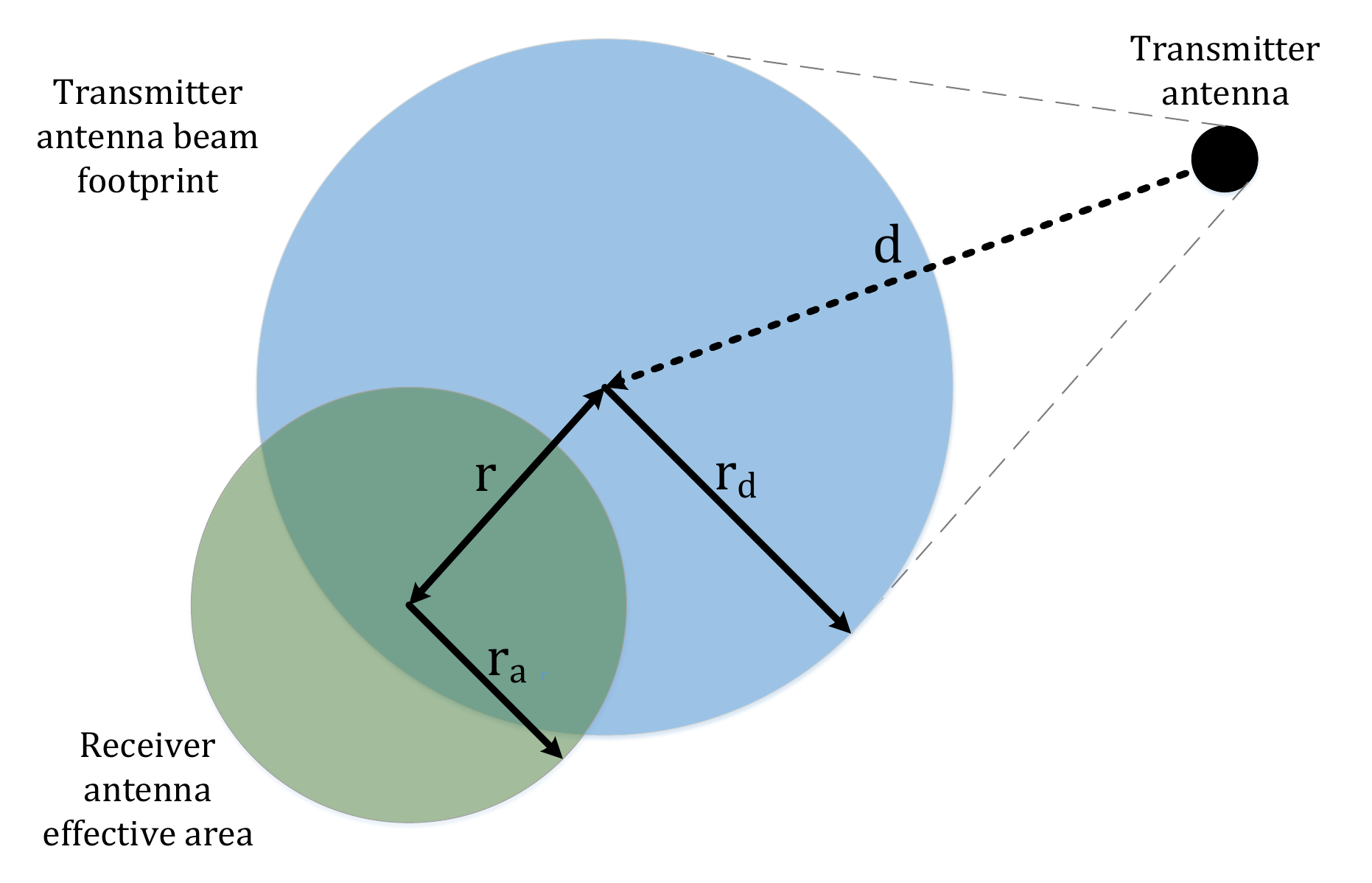}
    \caption{\textcolor{black}{Illustration of a beam misalignment} for the receiver's effective area with radius $r_a$ and the transmitter's beam footprint at distance $d$ shown by $r_d$ with the presence of the pointing error $r$.}
    \label{fig:misalignment}
\end{figure}

\textcolor{black}{It should be noted that \ac{THz} antennas create pencil sharp beams. Also, the motion of \ac{LEO} satellites with high velocity around $28\times10^{3}$ kph. Thus, the possible misalignment fading should be considered in \ac{THz}-empowered \acp{ISL}.} Under the circular beam assumption, considering beams with the radial distance, $r$, between their centers on the x-y plane, the misalignment coefficient $\zeta$ can be expressed as~\cite{farid_outage_2007}
\begin{align}
    \zeta(r ; d) \approx A_{o} \exp \left(-\frac{2 r^{2}}{w_{e q}^{2}}\right),
    \label{eq:zeta}
\end{align}
where
\begin{align}
A_{0}= & \left[\operatorname{erf}\left(\frac{\sqrt{\pi}r_{a}}{\sqrt{2}r_{d}}\right)\right]^{2}, \nonumber \\
w_{eq}^{2}= &r_{d}^{2} \frac{ \operatorname{erf}\left(\frac{\sqrt{\pi}r_{a}}{\sqrt{2}r_{d}}\right)}{ \frac{\sqrt{2}r_{a}}{r_{d}} \exp \left(-\frac{\pi r_{a}^2}{2r_{d}^{2}}\right)}.
\end{align}
$\operatorname{erf}(\cdot)$ is the error function. $A_{0}$, $w_{eq}$, and $r_{d}$ denote the collected power fraction in the aligned case (i.e., $r = 0$), equivalent beam width, and the beam waist at distance $d$, respectively. $r_a$ is the radius of the receiver antenna's effective area, which under the circular area assumption can be expressed by modifying the antenna aperture area equation~\cite{balanis2016antenna} as follows:
\begin{align}
    A_{e} = \pi r_a^2 = \frac{\lambda^2}{4\pi}G,
\end{align}
then,
\begin{align}
    r_a =  \frac{\lambda}{2\pi}\sqrt{G},
\end{align}
where $G$ denotes the antenna gain. Due to the fact that the displacement in both directions follows independent identical Gaussian distribution, the radial distance, $r$, can be \textcolor{black}{modeled by the following Rayleigh distribution:}
\begin{align}
    f_{r}(r)=\frac{r}{\sigma_{s}^{2}} \exp \left(-\frac{r^{2}}{2 \sigma_{s}^{2}}\right), \quad r>0,
    \label{eq:rayleigh}
\end{align}
where $\sigma_{s}^{2}$ stands for the receiver's jitter variance. By jointly utilizing \EQ{eq:zeta} and \EQ{eq:rayleigh}, the misalignment fading is introduced as follows:
\begin{align}
    f_{\zeta}(y) = \frac{\kappa^2}{A_{0}^{\kappa^2}}y^{\kappa^2-1}, \, 0\leq y\leq A_{0},
    \label{eq:misalignment_distribution}
\end{align}
where $\kappa = \frac{w_{eq}^2}{2 \sigma_{s}^2}$~\cite{boulogeorgos_error_2020}. Thus, the jitter variance appears as an important parameter affecting communication performance.

\begin{figure}[!t]
    \centering
    \includegraphics[width=\linewidth, page = 3]{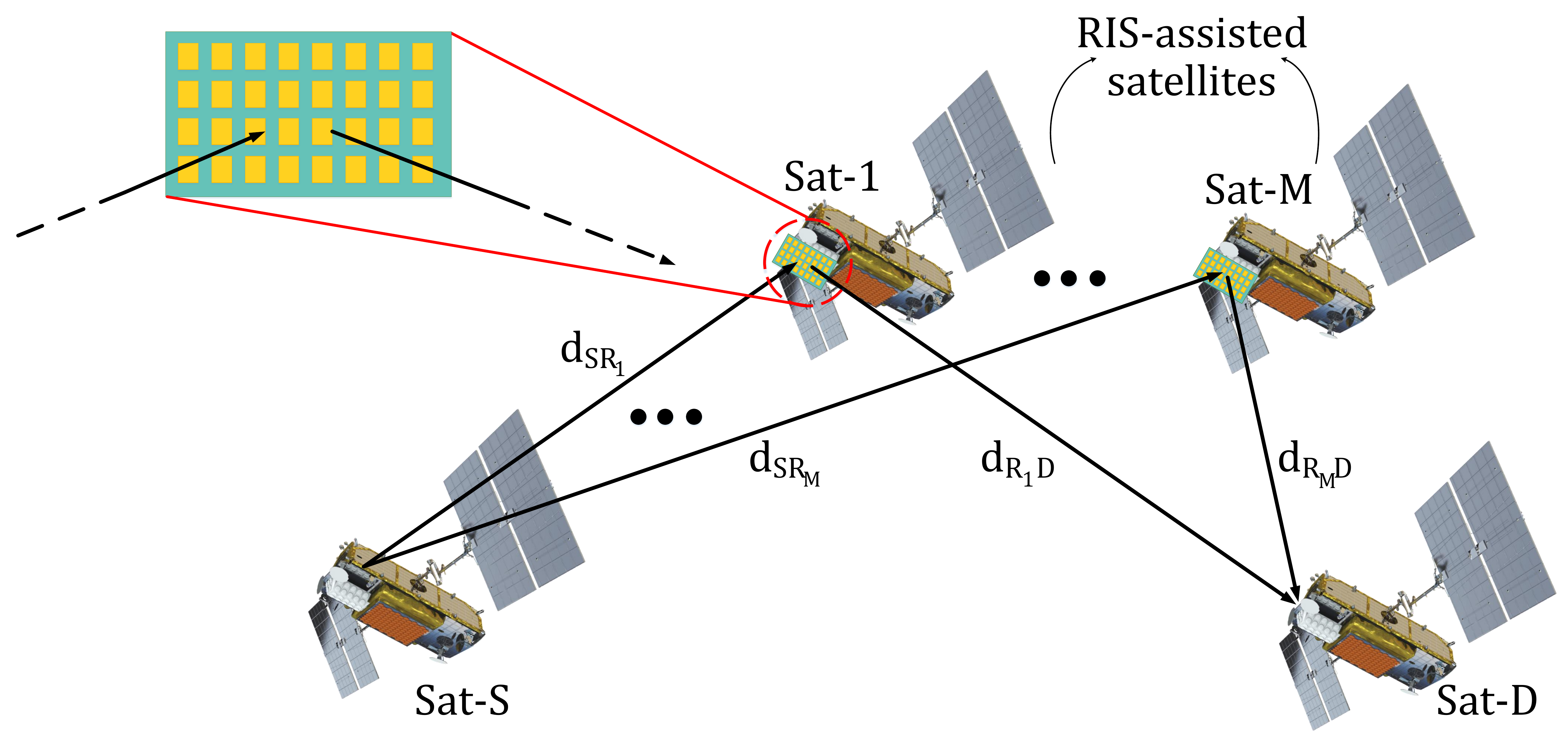}
    \caption{\textcolor{black}{Illustration of a simultaneous multiple RIS-assisted} satellite communication framework consisting of a source (Sat-S), destination (Sat-D), and $M$ satellites equipped with \ac{RIS}.}
    \label{fig:multiple_ris_1}
\end{figure}

\subsection{Single RIS-Assisted Wireless Channel and Associated Performance Analysis}\label{sec:single_ris_ch_model}

Under the flat fading assumption, the signal received by the destination node for $N$-element \ac{RIS} is expressed as follows:
\begin{align}
r=\sqrt{P_{t}P_{L}}\left(\sum_{i=1}^{N} \rho_{i}^{SR} e^{j \phi_{i}} \rho_{i}^{RD}\zeta_{i}^{SR}\zeta_{i}^{RD}\right) x+n,
\label{eq:received_signal}
\end{align}
where $\rho_{i}^{SR}$ and $\rho_{i}^{RD}$ denote the channel coefficients. The phase shift adjusted by \textcolor{black}{the} $i$th element of \ac{RIS} is represented by $\phi_{i}$. $x$ is the data symbol from possible constellation of an $M$-ary \ac{PSK} or \ac{QAM}. $n$ is zero-mean complex \ac{AWGN} with variance of $N_{0}$. $\phi_{i}$ \textcolor{black}{refers to} the adjustable phase shifts created by the $i^{th}$-elements of \ac{RIS}. $\zeta_{i}^{SR}$ and $\zeta_{i}^{RD}$ stand for the misalignment fading in the links source-to-\ac{RIS} and \ac{RIS}-to-destination. Also, $P_{t}$ and $P_L$ \textcolor{black}{refer to} the transmit power of source and the total path loss through the path propagating by the signal, respectively. The total path length is $d_{SR}^2 + d_{RD}^2$ in the near-field behavior; however, it becomes $d_{SR}^2 \times d_{RD}^2$ under the far-field of \ac{RIS}~\cite{ellingson2019path}. The path loss in the free-space is expressed as \textcolor{black}{follows:}
\begin{align}
    P_{L} =\left(\frac{\lambda}{4 \pi}\right)^{4} \frac{G_{i} G_{r}}{d_{SR}^{2} d_{RD}^{2}} \epsilon_{p},
\end{align}
where $G_{i}$ and $G_{r}$ \textcolor{black}{refer to} gains in the direction of the incoming and receiving waves, which are chosen as 30 dBi for 350 GHz operating frequency as given in~\cite{piesiewicz_performance_2008}. $\lambda$ denotes the wavelength corresponding to propagation frequency of wave, $\epsilon_{p}$ stands for the efficiency of \ac{RIS}. For the lossless \ac{RIS}, $\epsilon_{p}$ is equal to $1$. 

By rewriting the channel impulse responses in terms of channel amplitudes and phases as $\rho_{i}^{SR} = \alpha_{i}^{SR}e^{-j\beta_{i}^{SR}}$ and $\rho_{i}^{RD} = \alpha_{i}^{RD}e^{-j\beta_{i}^{RD}}$, Eq.~\EQ{eq:received_signal} can be expressed as follows:
\begin{align}
r=\sqrt{P_{t}P_{L}}\left(\sum_{i=1}^{N} \alpha_{i}^{SR}\alpha_{i}^{RD}e^{j(\phi_{i}-\beta_{i}^{SR}-\beta_{i}^{RD})}\zeta_{i}^{SR}\zeta_{i}^{RD}\right) x+n.
\label{eq:received_signal}
\end{align}
Then, the instantaneous \ac{SNR} observed at the receiver becomes
\begin{align}
\gamma = \frac{P_{t}\left|\sqrt{P_{L}}\left(\sum_{i=1}^{N} \alpha_{i}^{SR}\alpha_{i}^{RD}e^{j\psi}\zeta_{i}^{SR}\zeta_{i}^{RD}\right)\right|^2}{N_{0}},
\end{align}
where $\psi = \phi_{i}-\beta_{i}^{SR}-\beta_{i}^{RD}$ and it is zero for the maximum instantaneous \ac{SNR} as
\begin{align}
\gamma_{\max} = \frac{P_{t}\left|\sqrt{P_{L}}\left(\sum_{i=1}^{N} \alpha_{i}^{SR}\alpha_{i}^{RD}\zeta_{i}^{SR}\zeta_{i}^{RD}\right)\right|^2}{N_{0}} = \frac{A^2 P_{t}}{N_{0}},
\label{eq:max_snr}
\end{align} 
where $\alpha_{i}^{SR}$ and $\alpha_{i}^{RD}$ follow the Rician distribution which \textcolor{black}{may be} observed in the space communication due to the presence of solar scintillation~\cite{morabito2003solar} or a high power \ac{LOS} link with \ac{NLOS} links reflected from other space things. As described in~\cite{shaft1974relationship}, solar scintillation can be modeled with Rician fading. $\zeta_{i}^{SR}$ and $\zeta_{i}^{RD}$ are the misalignment fading coefficients following the distribution given in Eq.~\EQ{eq:misalignment_distribution}. According to the \ac{CLT}, $A$ follows \textcolor{black}{a} Gaussian distribution for $N \gg 1$ with the mean and variance given in \EQ{eq:mean_var}, where $L_{1/2}(\cdot)$ stands for the Laguerre polynomial of degree $1/2$. As $A$ follows \textcolor{black}{a} Gaussian distribution, \ac{SNR} is non-central chi-square distributed with one degree of freedom. To evaluate the \ac{BEP}, the \ac{MGF} of \textcolor{black}{a} non-central chi-square distribution~\cite{jg2001digital} is utilized as given in \EQ{eq:mgf}. By utilizing the \ac{SEP} expression for $M$-PSK signals given in~\cite{simon2005digital}, the error probability for \ac{BPSK} is obtained as in \EQ{eq:ber}. \textcolor{black}{This means} that the error probability degrades inversely $N^2$ while it proportionally increases with $(\sigma_s^2)^4$ due to the term $\kappa$ in the equation.

\begin{figure}[!t]
    \centering
    \includegraphics[width=\linewidth, page = 8]{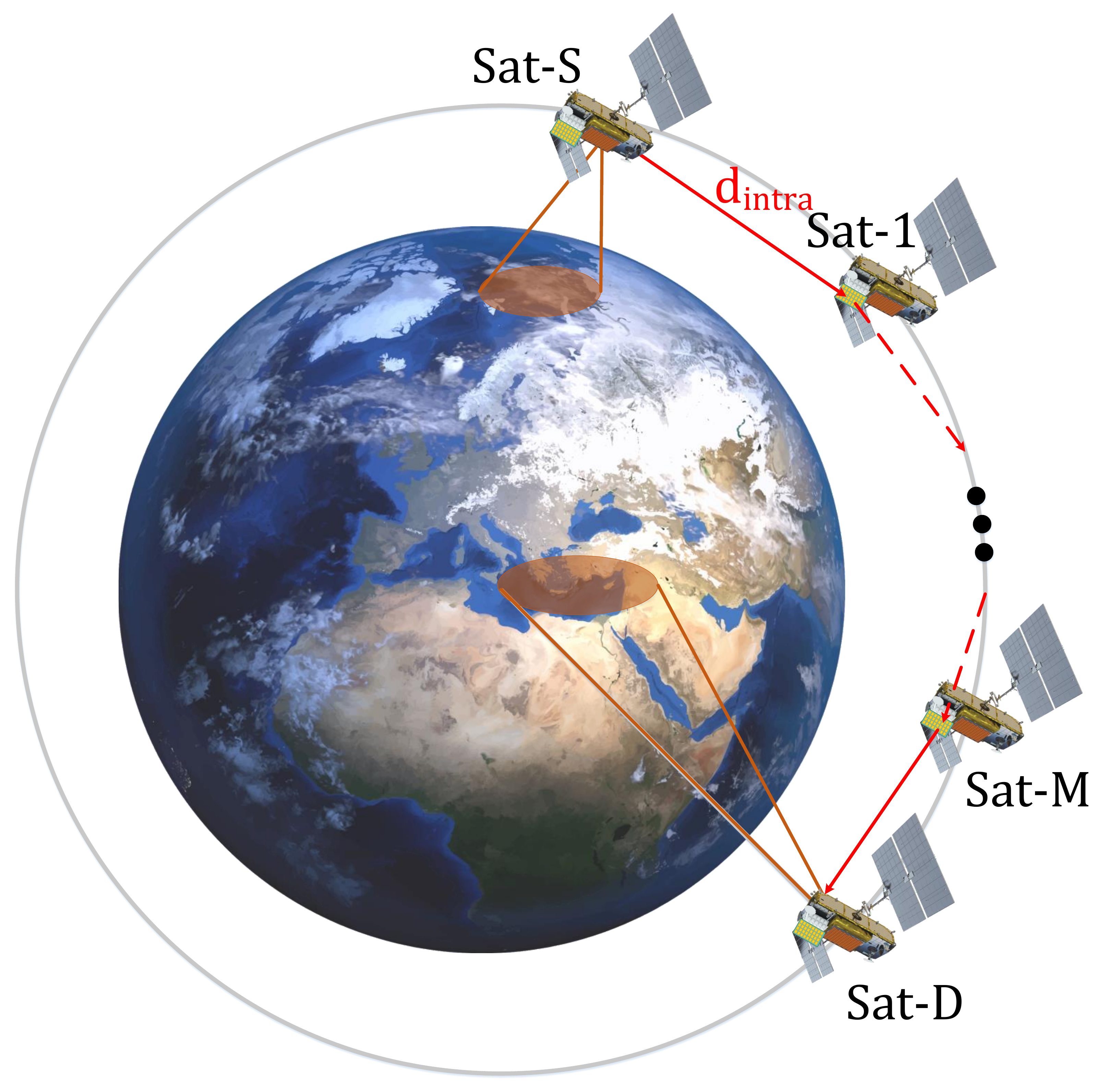}
    \caption{\textcolor{black}{Illustration of a} consecutive multiple RIS-assisted satellite communication framework consisting of a source (Sat-S), destination (Sat-D), and $M$ satellites equipped with \ac{RIS}, which are in the same orbit.}
    \label{fig:multiple_ris_2}
\end{figure}

\newcounter{mytempeqncnt20}
\begin{figure*}[ht!]
\normalsize
\setcounter{mytempeqncnt20}{\value{equation}}
\setcounter{equation}{17}
\begin{align}
      \E\left[A\right]= &\frac{N\sqrt{P_L}}{2\left(1+K\right)}\frac{\pi}{2}\left(L_{1/2}\left(-K\right)\right)^2\frac{\kappa^4}{\left(\kappa^2+1\right)^2}A_0^2, \nonumber\\
      \Var\left[A\right] = & N P_L\left(A_0^4\frac{\kappa^4}{\left(\kappa^2+2\right)^2}-\left[\frac{\pi}{4\left(1+K\right)}\left(L_{1/2}\left(-K\right)\right)^2\frac{\kappa^4}{\left(\kappa^2+1\right)^2}A_0^2\right]^2\right)
      \label{eq:mean_var}
\end{align}
\setcounter{equation}{\value{mytempeqncnt20}}
\hrulefill
\vspace*{1pt}
\end{figure*}
\setcounter{equation}{18}

\newcounter{mytempeqncnt0}
\begin{figure*}[ht!]
\normalsize
\setcounter{mytempeqncnt0}{\value{equation}}
\setcounter{equation}{18}
\begin{equation}
      M_{\gamma_{max}}(s) = \frac{\exp \left(-\frac{ \frac{p_t}{N_{0}} \frac{N^{2} P_{L} \pi^{2}}{16(1+K)^{2}}\left(L_{1 / 2}(-K)\right)^{4} \frac{\kappa^{8}}{\left(\kappa^{2}+1\right)^{4}} A_{0}^{4}}{1-2 s \frac{P t}{N_{0}} N P_{L}\left(\frac{A_{0}^{4} \kappa^{4}}{\left(\sigma^{2}+2\right) 2}-\left[\frac{\pi}{4(1+K)}\left(L_{1 / 2}(-K)\right)^{2} \frac{\kappa^{4}}{\left(\kappa^{2}+1\right)^{2}} A_{0}^{2}\right]^{2}\right)}\right)}{\sqrt{1-2 s\frac{P_t}{N_0} N P_{L}\left(\frac{A_{0}^{4} \kappa^{4}}{\left(\kappa^{2}+2\right)^{2}}-\left[\frac{\pi}{4(K+1)}\left(L_{1 / 2}(-K)\right)^{2} \frac{\kappa^{4}}{\left(\kappa^{2}+1\right)^{2}} A_{0}^{2}\right]^{2}\right)}}.
      \label{eq:mgf}
\end{equation}
\setcounter{equation}{\value{mytempeqncnt0}}
\hrulefill
\vspace*{1pt}
\end{figure*}
\setcounter{equation}{19}

\newcounter{mytempeqncnt100}
\begin{figure*}[ht!]
\normalsize
\setcounter{mytempeqncnt100}{\value{equation}}
\setcounter{equation}{19}
\begin{equation}
      P_{e} = \frac{1}{\pi}\int_{0}^{\pi/2} \frac{\exp \left(-\frac{ \frac{p_t}{N_{0}} \frac{N^{2} P_{L} \pi^{2}}{16(1+K)^{2}}\left(L_{1 / 2}(-K)\right)^{4} \frac{\kappa^{8}}{\left(\kappa^{2}+1\right)^{4}} A_{0}^{4}}{1+2 \sin^2(\omega)  \frac{P t}{N_{0}} N P_{L}\left(\frac{A_{0}^{4} \kappa^{4}}{\left(\sigma^{2}+2\right) 2}-\left[\frac{\pi}{4(1+K)}\left(L_{1 / 2}(-K)\right)^{2} \frac{\kappa^{4}}{\left(\kappa^{2}+1\right)^{2}} A_{0}^{2}\right]^{2}\right)}\right)}{\sqrt{1+2 \sin^2(\omega)\frac{P_t}{N_0} N P_{L}\left(\frac{A_{0}^{4} \kappa^{4}}{\left(\kappa^{2}+2\right)^{2}}-\left[\frac{\pi}{4(K+1)}\left(L_{1 / 2}(-K)\right)^{2} \frac{\kappa^{4}}{\left(\kappa^{2}+1\right)^{2}} A_{0}^{2}\right]^{2}\right)}}d\omega.
      \label{eq:ber}
\end{equation}
\setcounter{equation}{\value{mytempeqncnt100}}
\hrulefill
\vspace*{1pt}
\end{figure*}
\setcounter{equation}{20}

\subsection{Multiple RIS-Assisted Wireless Channel and Associated Performance Analysis}\label{sec:multi_ris_ch_model}
In this section, we investigate the performance \textcolor{black}{of} the multiple \ac{RIS}-assisted satellite networks. First, \textcolor{black}{like in}~\cite{yildirim2019propagation}, we derive a mathematical expression of the error probability for a simultaneous transmission over $M$ independent \acp{RIS}. Then, we analyze the error performance for the transmission path consisting \textcolor{black}{of} $M$ reflections over consecutive \acp{RIS}.

\subsubsection{Simultaneous Transmission over $M$ Independent \acp{RIS}}\label{sec:multi_ris_1}

For the scenario depicted in \FGR{fig:multiple_ris_1}, the received signal can be expressed as \textcolor{black}{follows:}
\begin{align}
r=\sqrt{P_{t}}\left[\sum_{k=1}^{M}\sqrt{P_{Lk}}\left(\sum_{i=1}^{N_k} \alpha_{i}^{SR_k}\alpha_{i,k}^{R_kD}e^{j\psi_k}\zeta_{i,k}^{SR_k}\zeta_{i,k}^{R_kD}\right)\right] x+n,
\label{eq:21}
\end{align}
where each element of Eq.~\EQ{eq:21} is the same as in Eq.~\EQ{eq:received_signal} for the k-th \ac{RIS}. Similar \textcolor{black}{to} Eq.~\EQ{eq:max_snr}, the maximum instantaneous \ac{SNR} is given as \textcolor{black}{follows:}
\begin{align}
\gamma_{\max} =& \frac{P_{t}\left|\sum_{k=1}^{M}\left[\sqrt{P_{Lk}}\left(\sum_{i=1}^{N_k} \alpha_{i,k}^{SR_k}\alpha_{i,k}^{R_kD}\zeta_{i,k}^{SR_k}\zeta_{i,k}^{R_kD}\right)\right]\right|^2}{N_{0}} \nonumber \\
=& \frac{A^2 P_{t}}{N_{0}}.
\label{eq:max_snr_2}
\end{align} 
By taking \ac{CLT} into account for $N_k \gg 1, \forall k = 1, 2, \cdots, M$, it can be said that $A$ follows \textcolor{black}{a} Gaussian distribution with the mean and variance given as in Eq.~\EQ{eq:multiple_ris_1_mean_var}. Similar to the single \ac{RIS}-assisted communication performance analysis, we can derive the error probability as stated in Eq.~\EQ{eq:multiple_ris_1_pe}, which implies that the error performance increases in proportion to the square of the total reflective elements \textcolor{black}{as given in}~\cite{yildirim2019propagation}. 
\newcounter{mytempeqncnt40}
\begin{figure*}[ht!]
\normalsize
\setcounter{mytempeqncnt40}{\value{equation}}
\setcounter{equation}{22}
\begin{align}
      \E\left[A\right]= &\left(\sum_{k=1}^{M}N_k\sqrt{P_{Lk}}\right)\times\frac{\pi}{4\left(1+K\right)}\left(L_{1/2}\left(-K\right)\right)^2\frac{\kappa^4}{\left(\kappa^2+1\right)^2}A_0^2, \nonumber\\
      \Var\left[A\right] = & \left(\sum_{k=1}^{M}N_kP_{Lk}\right)\times\left(A_0^4\frac{\kappa^4}{\left(\kappa^2+2\right)^2}-\left[\frac{\pi}{4\left(1+K\right)}\left(L_{1/2}\left(-K\right)\right)^2\frac{\kappa^4}{\left(\kappa^2+1\right)^2}A_0^2\right]^2\right)
      \label{eq:multiple_ris_1_mean_var}
\end{align}
\setcounter{equation}{\value{mytempeqncnt40}}
\hrulefill
\vspace*{1pt}
\end{figure*}
\setcounter{equation}{23}

\newcounter{mytempeqncnt1000}
\begin{figure*}[ht!]
\normalsize
\setcounter{mytempeqncnt1000}{\value{equation}}
\setcounter{equation}{23}
\begin{equation}
      P_{e} = \frac{1}{\pi}\int_{0}^{\pi/2} \frac{\exp \left(-\frac{ \left(\sum_{k=1}^{M}N_k\sqrt{P_{Lk}}\right)^2\frac{p_t}{N_{0}}\frac{\pi^{2}}{16(1+K)^{2}}\left(L_{1 / 2}(-K)\right)^{4} \frac{\kappa^{8}}{\left(\kappa^{2}+1\right)^{4}} A_{0}^{4}}{1+2 \sin^2(\omega)  \frac{P t}{N_{0}} \left(\sum_{k=1}^{M}N_kP_{Lk}\right)\left(\frac{A_{0}^{4} \kappa^{4}}{\left(\sigma^{2}+2\right) 2}-\left[\frac{\pi}{4(1+K)}\left(L_{1 / 2}(-K)\right)^{2} \frac{\kappa^{4}}{\left(\kappa^{2}+1\right)^{2}} A_{0}^{2}\right]^{2}\right)}\right)}{\sqrt{1+2 \sin^2(\omega)\frac{P_t}{N_0}\left(\sum_{k=1}^{M}N_kP_{Lk}\right)\left(\frac{A_{0}^{4} \kappa^{4}}{\left(\kappa^{2}+2\right)^{2}}-\left[\frac{\pi}{4(K+1)}\left(L_{1 / 2}(-K)\right)^{2} \frac{\kappa^{4}}{\left(\kappa^{2}+1\right)^{2}} A_{0}^{2}\right]^{2}\right)}}d\omega.
      \label{eq:multiple_ris_1_pe}
\end{equation}
\setcounter{equation}{\value{mytempeqncnt1000}}
\hrulefill
\vspace*{1pt}
\end{figure*}
\setcounter{equation}{24}

\subsubsection{Transmission over $M$ Consecutive \ac{RIS}-Assisted Satellites}

The double-\ac{RIS} reflected transmission proposed in~\cite{yildirim2019propagation} is generalized for $M$ consecutive \ac{RIS} reflections as depicted in \FGR{fig:multiple_ris_2}. Considering the baseband equivalent received signal given in~\cite{yildirim2019propagation}, we can generalize the received signal passing through a path consisting \textcolor{black}{of} $M$ consecutive \ac{RIS} reflections as follows:
\begin{align}
r=\sqrt{P_{t}P_{L}}\left[\sum_{i_1=1}^{N}\sum_{i_2=1}^{N}\cdots\sum_{i_M=1}^{N} \alpha_{i_1...i_M}e^{j\psi}\right] x+n,
\end{align}
where $\alpha_{i_1...i_M}$ and $\psi$ denote the overall channel amplitude and overall channel phase along with the reconfiguration phases of \acp{RIS}. It should be \textcolor{black}{noted that} the transmitter and receiver are fully aligned in this case. Configuring $\psi = 0$, the maximum instantaneous \ac{SNR} is obtained as \textcolor{black}{follows:}
\begin{align}
    \gamma_{\max} =& \frac{P_{t}\left|\sqrt{P_{L}}\left[\sum_{i_1=1}^{N}\sum_{i_2=1}^{N}\cdots\sum_{i_M=1}^{N} \alpha_{i_1...i_M}\right]\right|^2}{N_{0}} \nonumber \\ =& \frac{A^2 p_t}{N_{0}}.
\end{align}
For $N\gg1$, $A$ follows \textcolor{black}{a} Gaussian distribution with the mean and variance given below
\begin{align}
      \E\left[A\right]= &\frac{N^{M}\sqrt{P_{L}\pi}}{\sqrt{4\left(1+K\right)}}L_{1/2}\left(-K\right), \nonumber\\
      \Var\left[A\right] = &  N^{M}P_{L}\left(1-\frac{\pi}{4\left(1+K\right)}\left(L_{1/2}\left(-K\right)\right)^2\right).
      \label{eq:multiple_ris_2_mean_var}
\end{align}
By applying \textcolor{black}{the} same procedure as in \SEC{sec:multi_ris_1}, the error probability is obtained as \textcolor{black}{follows:}
\begin{equation}
      P_{e} = \frac{1}{\pi}\int_{0}^{\pi/2} \frac{\exp \left(-\frac{\frac{N^{2M}P_{L}\pi}{4\left(1+K\right)L_{1/2}^2\left(-K\right)}}{1 + 2\frac{N^{M}P_{L}}{sin^2\left(\omega\right)}\left(1-\frac{\pi}{4\left(1+K\right)}L_{1/2}^2\left(-K\right)\right)}\right)}{\sqrt{1 + 2\frac{N^{M}P_{L}}{sin^2\left(\omega\right)}\left(1-\frac{\pi}{4\left(1+K\right)}L_{1/2}^2\left(-K\right)\right)}} d\omega.
      \label{eq:multiple_ris_2_pe}
\end{equation}

\section{Numerical Results and Discussion}\label{sec:results}
\textcolor{black}{In this section, we present numerical results to support the theoretical findings of \SEC{sec:system_model}}. It should be noted that \textcolor{black}{ for the simulations in this section, the operating frequency of the system has been set to $350$ GHz. The antenna gain is chosen as $30$ dBi as given in~\cite{piesiewicz_performance_2008}}. As mentioned \textcolor{black}{above}, the channel amplitude coefficients follow a Rician distribution with parameter $K = 10$. First of all, we investigate the error performance for single \ac{RIS}-assisted satellite constellations. Then, multiple \acp{RIS} are employed in the satellite networks, and the communication performance analyses are carried out regarding these scenarios.

\subsection{Single RIS-Assisted THz Inter-Satellite Links}\label{sec:general_results}

\begin{figure}[!t]
    \centering
    \includegraphics[width=\linewidth]{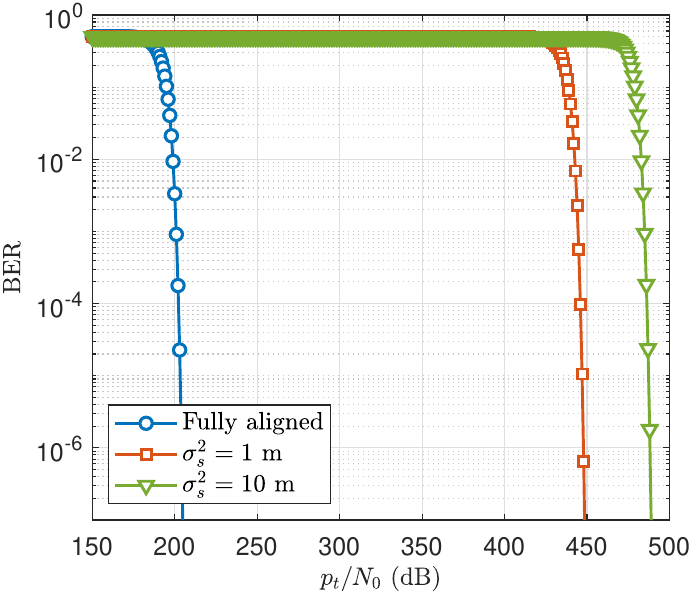}
    \caption{\ac{BER} performance of \textcolor{black}{a} single \ac{RIS}-assisted Starlink constellation with \ac{BPSK} signaling over $1024$-element \ac{RIS} for fully aligned transmission and transmission with misalignment jitter.}
    \label{fig:starlink_misalignment_comparison}
\end{figure}

First, \ac{THz} \acp{ISL} \textcolor{black}{with a single RIS are considered,} as depicted in \FGR{fig:system_model}. In this case, there is no direct link between the source and destination satellites, which is an acceptable assumption considering the satellite constellations~\cite{handley2018delay}. Due to fast relative \textcolor{black}{velocities} and possible beam tracking errors, misalignment fading has to be considered. Hence, in this section, misalignment fading is investigated for \ac{RIS}-assisted \ac{THz} satellite networks. To the best of \textcolor{black}{our} knowledge, this study \textcolor{black}{is the first to consider} the misalignment fading in \ac{RIS}-assisted \ac{THz} communications.

Two satellite constellations, Starlink and Iridium, are considered as examples, and numerical results are given \textcolor{black}{for} these satellite network constellations. Basically, the differences between these two constellations result from the number of orbits and satellites in each orbit. \textcolor{black}{These differences are tabulated in} \TAB{tab:distances}.

We assume that the distances Sat-S to Sat-R and Sat-R to Sat-D are equal to $d_{intra}$\textcolor{black}{; in other words, we can think of them being in the same orbit}. For Starlink, $d_{intra}$ is $945.4$ km while Iridium's intra-satellite distance is $4034$ km. 
\textcolor{black}{First}, the impact of misalignment fading on the error performance \textcolor{black}{is analyzed by comparing a} fully aligned system with some jitter variances. \FGR{fig:starlink_misalignment_comparison} reveals that in parallel with~\cite{boulogeorgos_analytical_2019}, alignment appears to be critical for \ac{THz} communication systems.

\begin{figure}[!t]
    \centering
    \includegraphics[width=\linewidth]{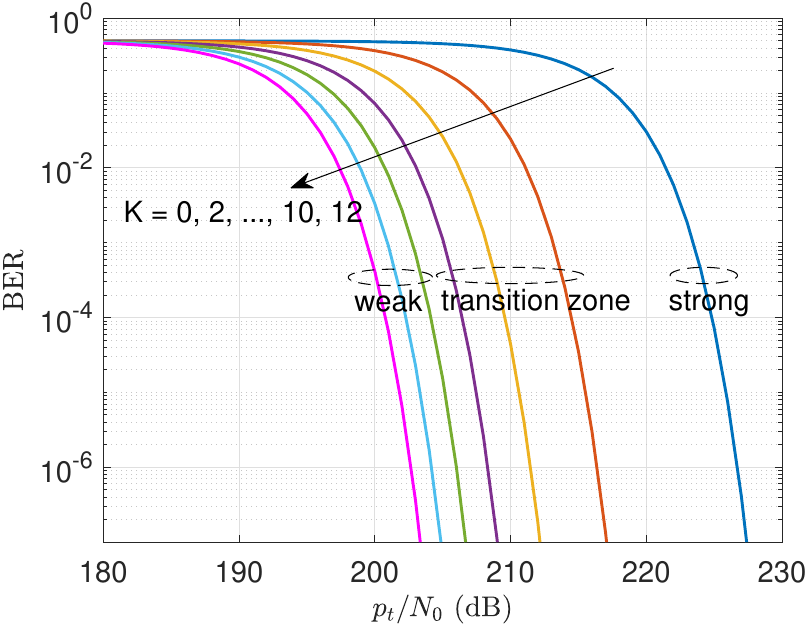}
    \caption{\ac{BER} performance of a single \ac{RIS}-assisted Starlink constellation with \ac{BPSK} signaling over $1024$-element \ac{RIS} \textcolor{black}{with a} changing solar scintillation which shows its effect as the Rician fading on the signal.}
    \label{fig:starlink_scintillation_comparison}
\end{figure}

Eq.~\EQ{eq:ber} shows that the jitter variance affects the probability of error in $\frac{\kappa^4}{\left(\kappa^2+1\right)^2}$ expression. Since there is a relationship similar to the reverse sigmoid function between the error probability and $\sigma_s$, $\frac{\kappa^4}{\left(\kappa^2+1\right)^2}$ converges to its minimum value for small values of $\sigma_s$. The beginning of this convergence starts around $\sigma_{s}^{2} = 1\,\mathrm{m}^2$. In this study, since $\sigma_{s}^{2}< 1\,\mathrm{m}^2$ values cause almost similar error probability, no analysis is included for \textcolor{black}{a} jitter variance less than $1\,\mathrm{m}^2$. An increase in the variance jitter degrades the error performance until the jitter variance reaches the point where $\frac{\kappa^4}{\left(\kappa^2+1\right)^2}$ converges its maximum. \textcolor{black}{Because of the long distances involved,} a jitter variance of up to $10\,\mathrm{m}^2$ is considered. Since a jitter variance greater \textcolor{black}{than this could make communication unfeasible,} it is not reasonable to evaluate it in this study.

In deep space channels, solar scintillation has to be taken into account \textcolor{black}{since} it results in fluctuations in the signal envelope. As aforementioned \textcolor{black}{above, it follows a Rician distribution}. Since the error rate is almost inversely proportional to the square of the shape parameter of Rician fading. The solar scintillation is investigated for three intensity regions: weak scintillation ($K\geqslant7$), transition zone ($7>K>0$), and strong scintillation ($K=0$). The impact of solar scintillation on the error performance is \textcolor{black}{shown} in \FGR{fig:starlink_scintillation_comparison}. Strong scintillation appears to significantly reduce communication performance. Although increased solar scintillation reduces error performance, \textcolor{black}{its effects are not as serious as misalignment fading.}

For varying number of \ac{RIS} elements, the error performances are depicted in \FGR{fig:single_ris_starlink} for the system resembling Starlink. \textcolor{black}{As we can see, when} the number of \ac{RIS} elements is doubled, $6$ dB power is gained to satisfy the same error performance as given in Eq.~\EQ{eq:ber}. Also, the misalignment fading is investigated in \FGR{fig:single_starlink_1} and \FGR{fig:single_starlink_10}. First, $\sigma_s^2$ is assumed to be $1$~m. Considering the results \textcolor{black}{shown} in both figures, when the jitter variance is increased $10$ times, $40$ dB more power must be used to ensure the same error performance as shown in Eq.~\EQ{eq:ber}. These results \textcolor{black}{indicate} that \ac{THz} satellite networks should either use \ac{RIS} with \textcolor{black}{a} higher number of elements or beam tracking methods that work with very low errors at the expense of high computational \textcolor{black}{costs} must be adopted. Similar behaviors are observed in \FGR{fig:single_ris_iridium} for Iridium with a higher path loss since it has longer intra-satellite distances than Starlink. As a result, more output power is needed to compensate for path loss.

\begin{figure}[!t]
    \centering
    \subfigure[]{
        \label{fig:single_starlink_1}
        \includegraphics[width=0.46\columnwidth]{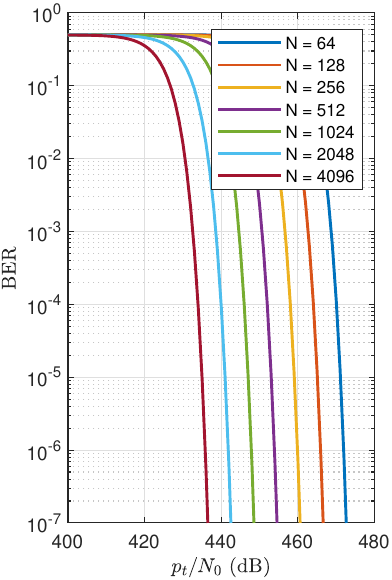}}
    \quad
    \subfigure[]{
        \label{fig:single_starlink_10}
        \includegraphics[width=0.46\columnwidth]{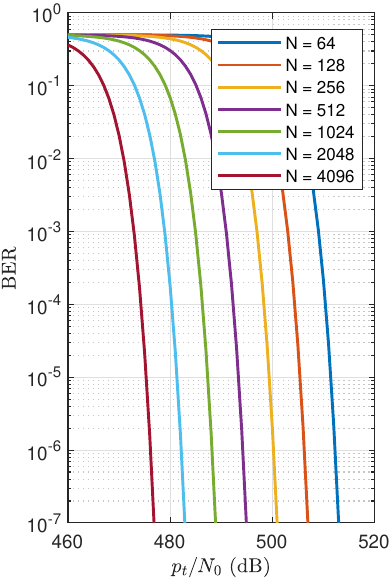}}
    \caption{\ac{BER} performance of \textcolor{black}{a} single \ac{RIS}-assisted Starlink for different \ac{RIS} sizes \textcolor{black}{with} \ac{BPSK} signaling and the misalignment fading with \textcolor{black}{the} receiver's jitter variance of (a) $\sigma_s^2 = 1\, \mathrm{m}^2$, (b) $\sigma_s^2 = 10\, \mathrm{m}^2$.}
    \label{fig:single_ris_starlink}
\end{figure}

\begin{figure}[!t]
    \centering
    \subfigure[]{
        \label{fig:single_iridium_1}
        \includegraphics[width=0.46\columnwidth]{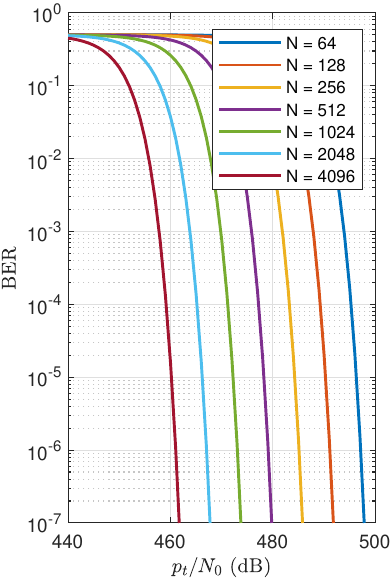}}
    \quad
    \subfigure[]{
        \label{fig:single_iridium_10}
        \includegraphics[width=0.46\columnwidth]{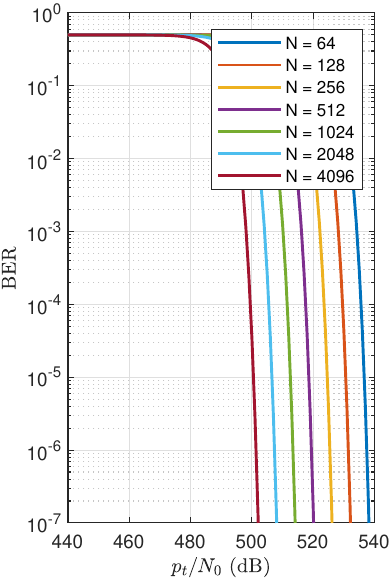}}
    \caption{\ac{BER} performance of \textcolor{black}{a} single \ac{RIS}-assisted Iridium for different \ac{RIS} sizes \textcolor{black}{with} \ac{BPSK} signaling and the misalignment fading with \textcolor{black}{the} receiver's jitter variance of (a) $\sigma_s^2 = 1\, \mathrm{m}^2$, (b) $\sigma_s^2 = 10\, \mathrm{m}^2$.}
    \label{fig:single_ris_iridium}
\end{figure}

\subsection{Multiple RIS-Assisted THz Inter-Satellite Links}\label{sec:multiple_results}
\textcolor{black}{Since satellite networks are designed to be densely interconnected, multiple satellites equipped with RISs can be deployed in a cooperative manner.} Thus, the diversity order can escalate. \textcolor{black}{To this end}, we investigate two different cases. \textcolor{black}{First, we look at simultaneous transmission over multiple satellites, followed by transmission over $M$ consecutive satellites.}

\subsubsection{Performance Analysis for Simultaneous Transmission over Multiple Satellites}

As \textcolor{black}{shown} in \FGR{fig:multiple_ris_1}, $M$ satellites equipped with \acp{RIS} can simultaneously transmit the incident signal to the destination satellite by increasing diversity order and dropping down the path loss exponent~\cite{basar_wireless_2019}. Eq.~\EQ{eq:multiple_ris_1_pe} clearly shows that the error probability is inversely proportional \textcolor{black}{to} the square of the total number of \ac{RIS} elements. It should be noted that in the light of Eq.~\EQ{eq:multiple_ris_1_pe}, the misalignment fading \textcolor{black}{remains the same as in a single RIS case.} In this scenario, $M$ is selected as $2$ with \textcolor{black}{a} varying number of \ac{RIS} elements. It is assumed that each distance is equal to $d_{intra}$; hence, $P_{Lk} = P_L,\, \forall k = 1, 2$. Also, the number of \ac{RIS} elements is selected as same. In \FGR{fig:simultaneous_multiple_ris_starlink} and \FGR{fig:simultaneous_multiple_ris_iridium}, the error performances are investigated for Starlink and Iridium, respectively. In this case, the error rate is proportional to $(2N)^{-2}$. As a result, the same error rate can be achieved with $6$ dB less power \textcolor{black}{than the} single \ac{RIS} case. 
 
 For the achievable rate analysis, $d_{SR_{1}}$ and $d_{R_{1}D}$ is set to $d_{intra}$. $d_{SR_{2}}$ and $d_{R_{2}D}$ vary from $d_{nearest}$ to $d_{farthest}$ to create a scenario such that Sat-S, Sat-1, and Sat-D are consecutive satellites in the same orbit, while Sat-2 is in the nearby orbit. Due to the fact that the positions of the satellites which are in different orbits continuously change and thus, the distance between them fluctuates during their orbits, we evaluate the capacity analysis for the upper and lower limits at the nearest and farthest distances. \FGR{fig:simultaneous_multiple_ris_capacity_starlink} and \FGR{fig:simultaneous_multiple_ris_capacity_iridium} \textcolor{black}{show} achievable rate for Starlink and Iridium satellites equipped with $1024$-element \acp{RIS} for the satellite position between two extreme distances. As expected, the maximum rate is achieved when the distances are minimum. A $10$-fold increase in misalignment fading causes the maximum achievable data rate to decrease by $5$ bits/s/Hz. For \textcolor{black}{the} \ac{THz} band, this decrease cannot be overlooked since the total data rate decrease reaches $300$ Gbps for $60$ GHz bandwidth. Owing to fact that the path loss between two consecutive satellites in Starlink is less than Iridium network, the achievable rate for Iridium is less. Considering the cost of construction and maintenance of satellites, increasing the number of satellites in each orbit \textcolor{black}{is not} feasible. \textcolor{black}{However}, increasing the number of \ac{RIS} elements on each satellite can provide cost-effective solution to reach the desired achievable rate.

\begin{figure}[!t]
    \centering
    \subfigure[]{
        \label{fig:simultaneous_multiple_ris_starlink_1}
        \includegraphics[width=0.46\columnwidth]{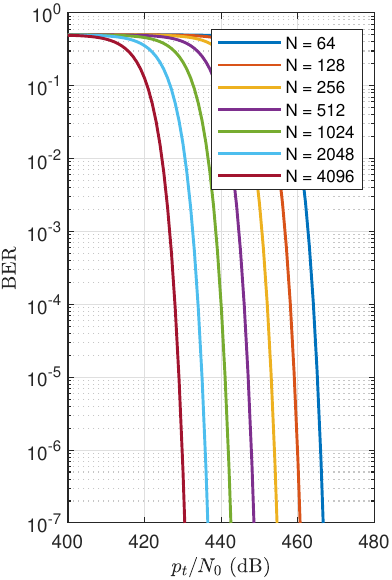}}
    \quad
    \subfigure[]{
        \label{fig:simultaneous_multiple_ris_starlink_10}
        \includegraphics[width=0.46\columnwidth]{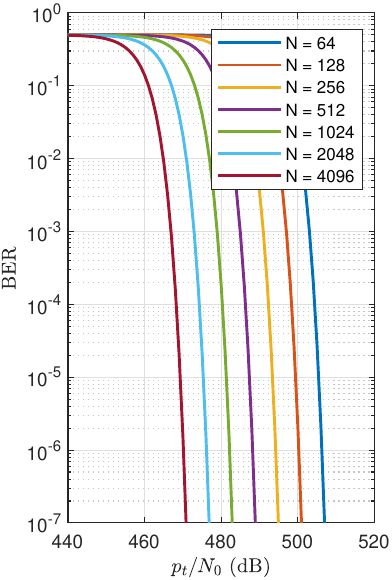}}
    \caption{\ac{BER} performance of \textcolor{black}{a} double simultaneous \ac{RIS}-assisted Starlink constellation for different \ac{RIS} sizes \textcolor{black}{with} \ac{BPSK} signaling and \textcolor{black}{a} misalignment fading with \textcolor{black}{the} receiver's jitter variance of (a) $\sigma_s^2 = 1\, \mathrm{m}^2$ and (b) $\sigma_s^2 = 10\, \mathrm{m}^2$. The distances are assumed as $d_{intra}$.}
    \label{fig:simultaneous_multiple_ris_starlink}
\end{figure}

\begin{figure}[!t]
    \centering
    \subfigure[]{
        \label{fig:simultaneous_multiple_ris_iridium_1}
        \includegraphics[width=0.46\columnwidth]{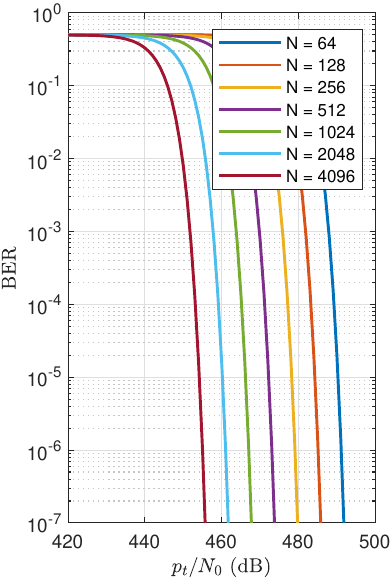}}
    \quad
    \subfigure[]{
        \label{fig:simultaneous_multiple_ris_iridium_10}
        \includegraphics[width=0.46\columnwidth]{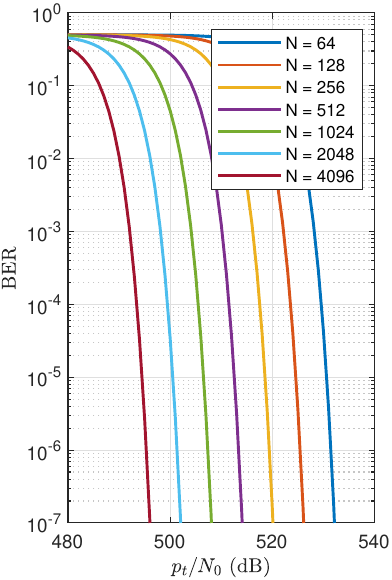}}
    \caption{\ac{BER} performance of \textcolor{black}{a} double simultaneous \ac{RIS}-assisted Iridium constellation for different \ac{RIS} sizes \textcolor{black}{with} \ac{BPSK} signaling and \textcolor{black}{a} misalignment fading with \textcolor{black}{the} receiver's jitter variance of (a) $\sigma_s^2 = 1\, \mathrm{m}^2$ and (b) $\sigma_s^2 = 10\, \mathrm{m}^2$. The distances are assumed as $d_{intra}$.}
    \label{fig:simultaneous_multiple_ris_iridium}
\end{figure}

\begin{figure*}[!t]
    \centering
    \subfigure[]{
        \label{fig:simultaneous_multiple_ris_capacity_starlink_1}
        \includegraphics[width=0.46\linewidth]{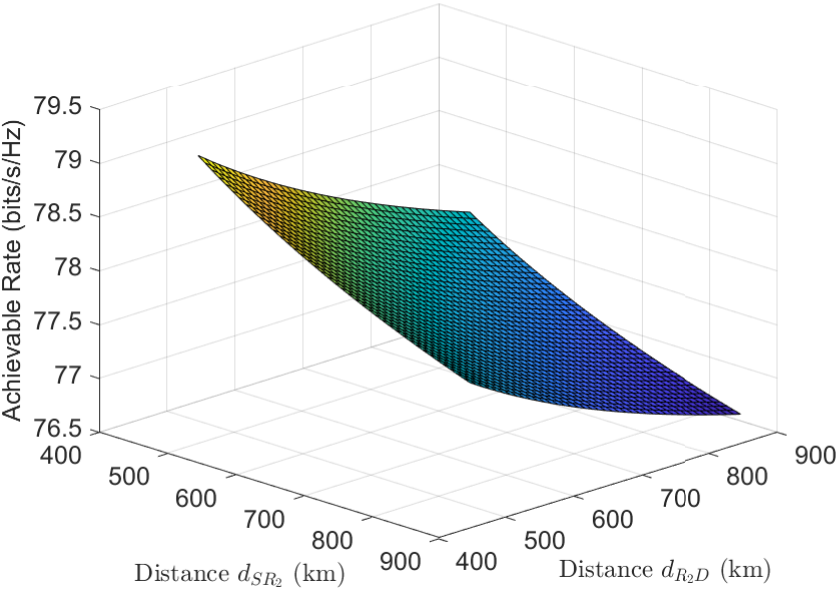}}
    \quad
    \subfigure[]{
        \label{fig:simultaneous_multiple_ris_capacity_starlink_10}
        \includegraphics[width=0.46\linewidth]{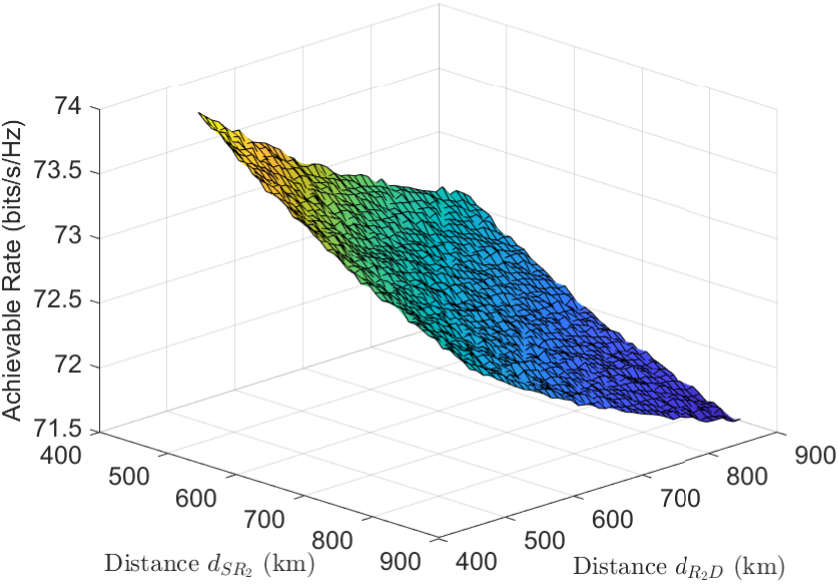}}
    \caption{Achievable data rate performance \textcolor{black}{for the} double simultaneous \acp{RIS}-assisted Starlink constellation for varying distances (i.e., $d_{SR_{2}}$ and $d_{R_{2}D}$) from $d_{nearest}$ to $d_{farthest}$ \textcolor{black}{with} two misalignment cases: (a) $\sigma_s^2 = 1\, \mathrm{m}^2$, (b) $\sigma_s^2 = 10\, \mathrm{m}^2$ \textcolor{black}{where} $N = 1024$. $d_{SR_{1}}$ and $d_{R_{1}D}$ are kept constant as $d_{intra}$.}
    \label{fig:simultaneous_multiple_ris_capacity_starlink}
\end{figure*}

\begin{figure*}[!t]
    \centering
    \subfigure[]{
        \label{fig:simultaneous_multiple_ris_capacity_iridium_1}
        \includegraphics[width=0.46\linewidth]{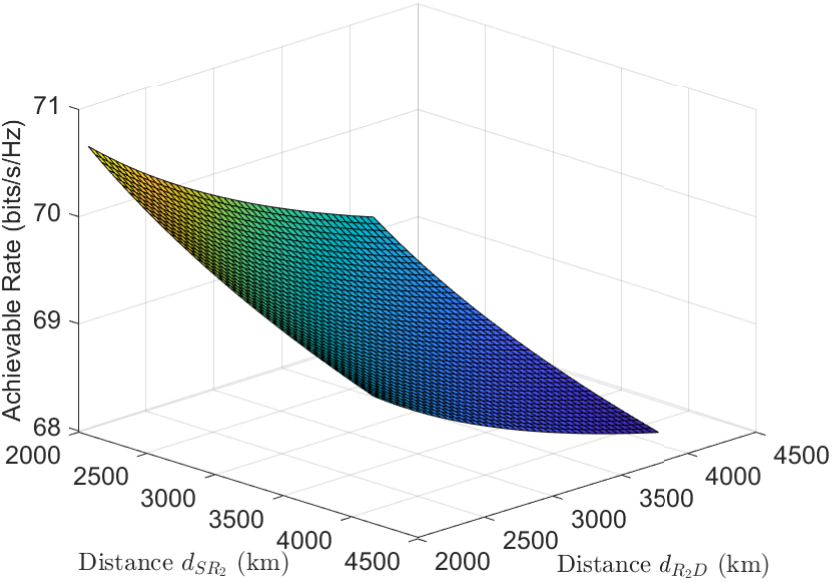}}
    \quad
    \subfigure[]{
        \label{fig:simultaneous_multiple_ris_capacity_iridium_10}
        \includegraphics[width=0.46\linewidth]{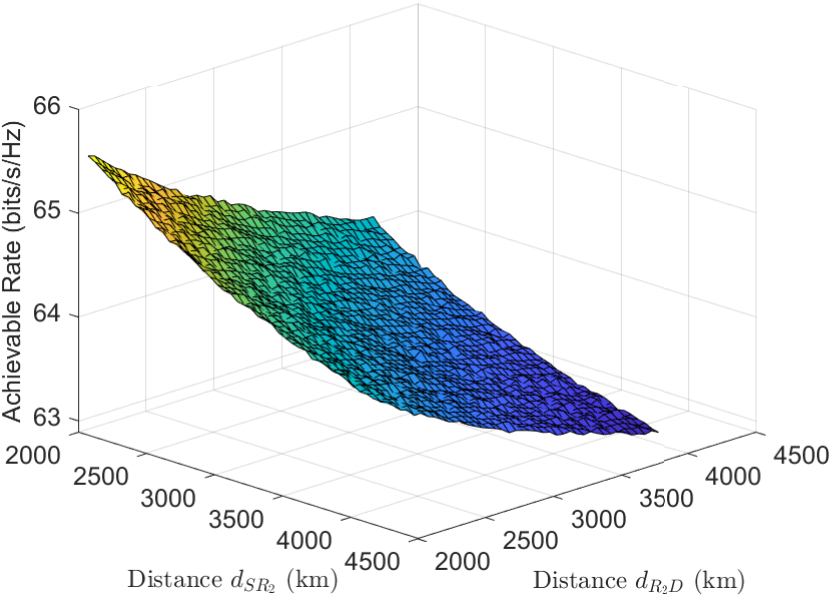}}
    \caption{Achievable data rate performance \textcolor{black}{for the} double simultaneous \acp{RIS}-assisted Iridium constellation for varying distances (i.e., $d_{SR_{2}}$ and $d_{R_{2}D}$) from $d_{nearest}$ to $d_{farthest}$ \textcolor{black}{with} two misalignment cases: (a) $\sigma_s^2 = 1\, \mathrm{m}^2$, (b) $\sigma_s^2 = 10\, \mathrm{m}^2$ \textcolor{black}{where} $N = 1024$. $d_{SR_{1}}$ and $d_{R_{1}D}$ are kept constant as $d_{intra}$.}
    \label{fig:simultaneous_multiple_ris_capacity_iridium}
\end{figure*}

\begin{figure*}[!t]
    \centering
    \subfigure[]{
        \label{fig:consecutive_multiple_ris_capacity_starlink}
        \includegraphics[width=0.46\linewidth]{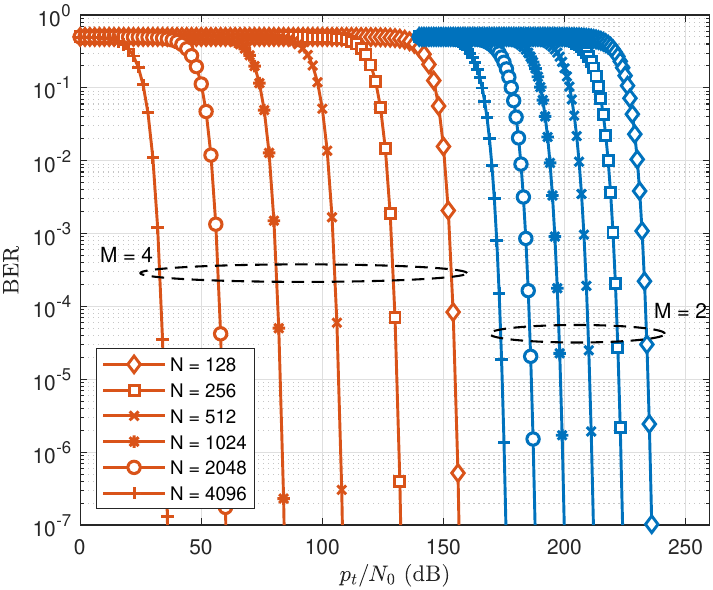}}
    \quad
    \subfigure[]{
        \label{fig:consecutive_multiple_ris_capacity_iridium}
        \includegraphics[width=0.46\linewidth]{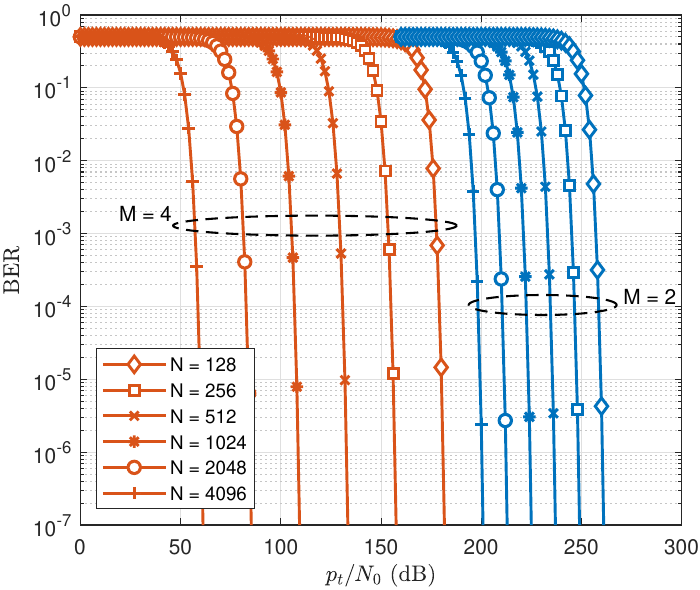}}
    \caption{\ac{BER} performance for the transmission over consecutive satellites equipped with $N$-element \ac{RIS} \textcolor{black}{in} two different constellations: (a) Starlink, (b) Iridium. \textcolor{black}{In this case, all satellites are assumed to be in the same orbit;} thus, the inter-satellite distance is $d_{intra}$.}
    \label{fig:consecutive_multiple_ris_capacity}
\end{figure*}

\subsubsection{Performance Analysis for Transmission over $M$ Consecutive Satellites}

Finally, the transmission sequentially through satellites scenario in the same orbit is investigated because satellite networks are designed to allow communication between different orbits in a limited way while allowing easy transmission on the same orbit~\cite{handley2018delay}. In this scenario, since all satellites are in the same orbit, the distance between the satellites is constant, so path losses between satellites can be considered equal. It should be \textcolor{black}{noted} that the fully aligned case is considered in this scenario.

As Eq.~\EQ{eq:multiple_ris_2_pe} embodies the relation $P_e \propto N^{2M}$, increasing \textcolor{black}{the} number of consecutive satellites significantly improves the error performance. For example, increasing from $M=2$ to $M=4$ results in approximately $120$ dB power gain when $1024$-element \acp{RIS} are employed in satellite networks. \textcolor{black}{The results can be clearly seen in \FGR{fig:consecutive_multiple_ris_capacity}. As we can see there, when} the number of \ac{RIS} elements is doubled, the power gain is $12$ dB and $24$ dB for $M = 2$ and $M = 4$, respectively. These results indicate that the power gain in transmission along the orbit can increase significantly.

\section{Open Issues}\label{sec:open_issues}
\textcolor{black}{We can observe from the foregoing that} satellite-mounted \ac{RIS} communication systems can increase error probability performance, but there are some open issues. Due to the continuous movement of satellites \textcolor{black}{along} their orbits, their positions relative to each other change. \textcolor{black}{Therefore, adaptive and steerable beam antenna designs are clearly needed.} In addition, since the orbits of the satellites can be planned \textcolor{black}{and} predicted, coarse beam alignment can be performed by utilizing a look-up table of the positions of the other satellites in each satellite. Then, beam steering methods can be used for \textcolor{black}{precise} alignment. Furthermore, the derivations in this study show that the system can be less affected by misalignment loss by using antennas with large beamwidths. \textcolor{black}{As discussed in~\cite{luo2019graphene, wu2018large, gomez2019low}, it is possible to produce antennas} with large beamwidths in the \ac{THz} \textcolor{black}{region by using semiconductors such as graphene and manufacturing techniques such as micro-machining.} \textcolor{black}{Finally, a deeper investigation of solar scintillation in the THz band is needed to improve the accuracy of the performance analysis.}

Since the potential of the \ac{THz} region has only \textcolor{black}{just begun to be explored} in terms of communication, \ac{THz} hardware has \textcolor{black}{yet to reach the desired} maturity level. \textcolor{black}{In particular,} there is a serious need for amplifier designs for inter-satellite communication links. \textcolor{black}{Still it should be noted that the RIS-supported communication} proposed in this study can be used in the lower \textcolor{black}{bands for satellite-to-ground} and inter-satellite connections. The derivations indicate that halving the operation frequency can reduce the required power by $12$ dB at the cost of a decrease in the achievable rate owing to the presence of the narrowband at lower frequency regions.

Furthermore, a path that can minimize the target performance metrics, such as round trip time, can be investigated by using \acp{RIS} with an optimization framework. Thus, \acp{RIS} can be used not only to improve communication performance but also to address other requirements, such as reducing the latency between two nodes. By modeling this problem as \textcolor{black}{a} traveling sales man problem, the path satisfying the minimum latency can be solved by mixed integer nonlinear programming. Furthermore, a performance analysis is \textcolor{black}{needed for when satellites move through their orbits. This can be obtained by utilizing orbit modeling tools,} such as SaVi.

\section{Conclusions}\label{sec:conclusion}

In this study, a solution to both frequency scarcity and low-complexity system design \textcolor{black}{was} proposed by the joint use of \ac{RIS} and \ac{THz} band in \ac{LEO} \acp{ISL}. \textcolor{black}{It was shown to be possible to increase achievable rates with the ultra-wide bandwidth provided by} the \ac{THz} wave and the \ac{RIS} acting like N-elements virtual \ac{MIMO} system. The derived mathematical expressions and simulation results \textcolor{black}{demonstrated} that an increase in the number of \acp{RIS} can reduce the required transmission power targeting the same error probability. It \textcolor{black}{was also shown} that error probability decreases inversely proportional to the square of the number of elements of \ac{RIS}. \textcolor{black}{Furthermore, we considered the issue of misalignment fading, which can occur in THz LEO satellite networks and which were validated with simulation results.} \textcolor{black}{Our} study shows that misalignment between the antennas gives rise to a severe drop in the received power. This indicates that a beam tracker is needed to ensure a convenient alignment system between the antennas. Also, it is shown by both derivations and simulation results that solar scintillation has a significant effect on error performance. In addition, considering the node density of the next-generation LEO satellites networks, cooperative communication techniques \textcolor{black}{has emerged} as a promising tool to improve the 
\ac{ISL} performances. \textcolor{black}{Finally, we investigated the performance of simultaneous and consecutive transmission between satellites equipped with more than one RIS, which demonstrated that simultaneous and consecutive transmission enhances communication in terms of error probability.} Moreover, the achievable rate \textcolor{black}{was} simulated for the position of the satellites in space under both the aligned case and misalignment fading. 

\bibliographystyle{IEEEtran}
\bibliography{main}

%%%%%%%%%%%%%%%% BIOGRAPHIES %%%%%%%%%%%%%%%%%%%%%%
\begin{IEEEbiography}
    [{\includegraphics[width=1in,height=1.25in,clip,keepaspectratio]{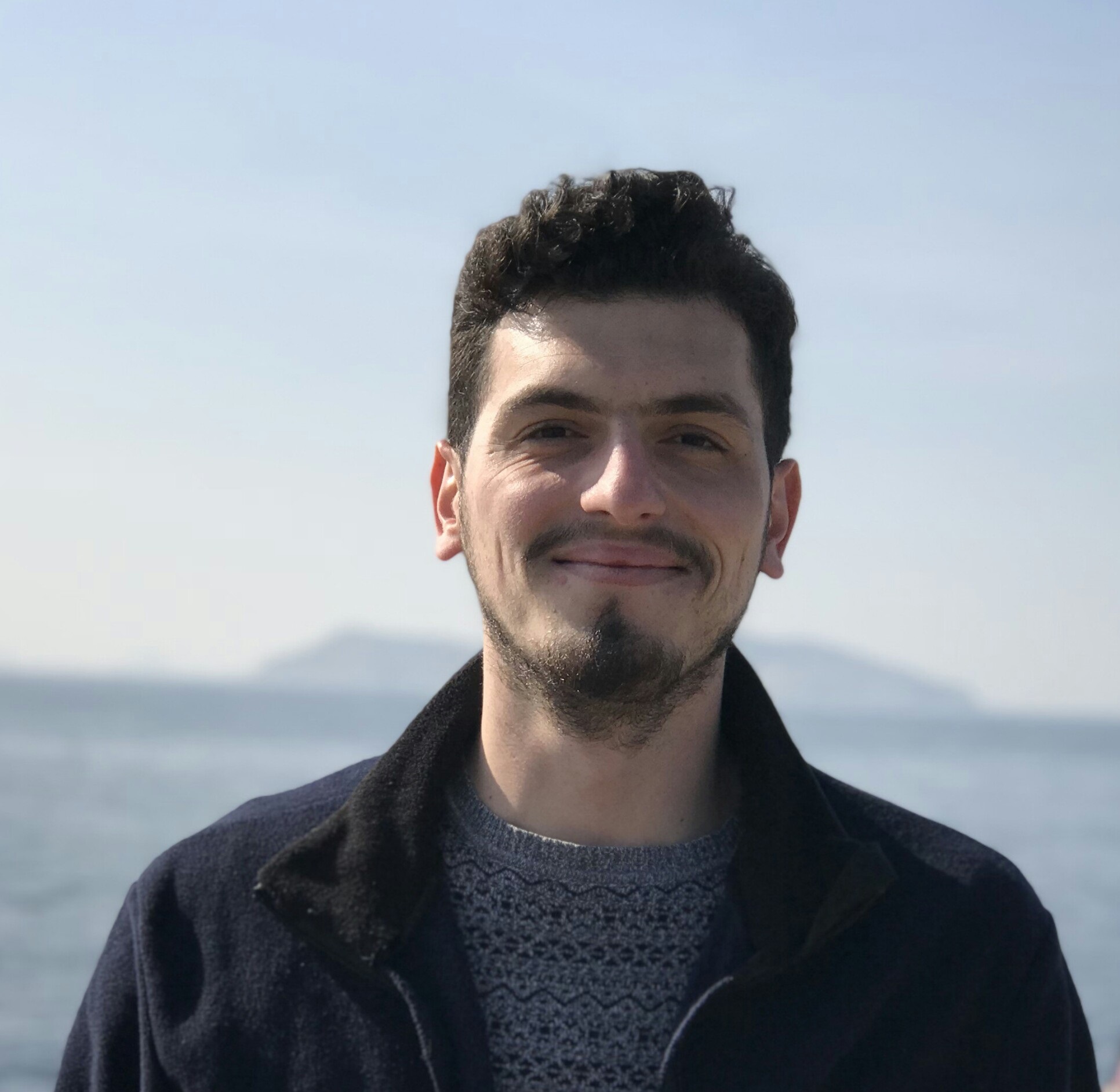}}]{K{\"{u}}r{\c{s}}at Tekb{\i}y{\i}k} [StM'19] (tekbiyik@itu.edu.tr) received his B.Sc. and M.Sc. degrees (with high honors) in electronics and communication engineering from Istanbul Technical University, Istanbul, Turkey, in 2017 and 2019, respectively. He is currently pursuing his Ph.D. degree in telecommunications engineering in Istanbul Technical University. His research interests include algorithm design for signal intelligence, next-generation wireless communication systems, terahertz wireless communications, and machine learning.
\end{IEEEbiography}

\begin{IEEEbiography}
    [{\includegraphics[width=1in,height=1.25in,clip,keepaspectratio]{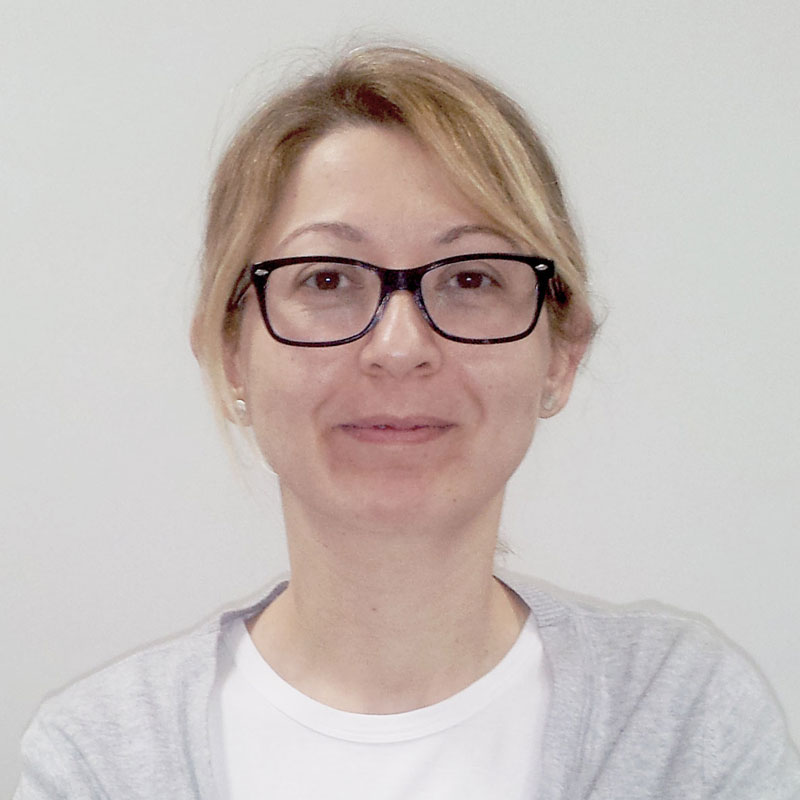}}]{G{\"{u}}ne{\c{s}} Karabulut Kurt} [StM'00, M'06, SM'15] (gkurt@itu.edu.tr) received the Ph.D. degree in electrical engineering from the University of Ottawa, Ottawa, ON, Canada, in 2006. Between 2005 and 2008, she was with TenXc Wireless, and Edgewater Computer Systems, in Ottawa Canada. From 2008 to 2010, she was with Turkcell R\&D Applied Research and Technology, Istanbul. Since 2010, she has been with Istanbul Technical University. She is also an Adjunct Research Professor at Carleton University. She is serving as an Associate Technical Editor of IEEE Communications Magazine. 
    
    %received the B.S. degree (with honors) in electronics and electrical engineering from Bogazici University, Istanbul, Turkey, in 2000 and the M.A.Sc. and Ph.D. degrees in electrical engineering from the University of Ottawa, Ottawa, ON, Canada, in 2002 and 2006, respectively. From 2000 to 2005, she was a Research Assistant with the CASP Group, University of Ottawa. Between 2005 and 2006, she was with TenXc Wireless, where she worked on location estimation and radio frequency identification systems. From 2006 to 2008, she was with Edgewater Computer Systems Inc., where she worked on high-bandwidth networking in aircraft and priority-based signaling methodologies. From 2008 to 2010, she was with Turkcell R\&D Applied Research and Technology, Istanbul. Since 2010, she has been with Istanbul Technical University, where she is currently a full Professor. Dr. Karabulut Kurt is a Marie Curie Fellow and her project REALMARS is selected to be a success story by European Commission Research Executive Agency. She is the author of numerous publications and international patents. Dr. Karabulut was involved in various research projects including STREP, EUREKA and ITEA-2 projects on wireless networks. She served a management committee member of the COST Action 1104.
\end{IEEEbiography}

\begin{IEEEbiography}
    [{\includegraphics[width=1in,height=1.25in,clip,keepaspectratio]{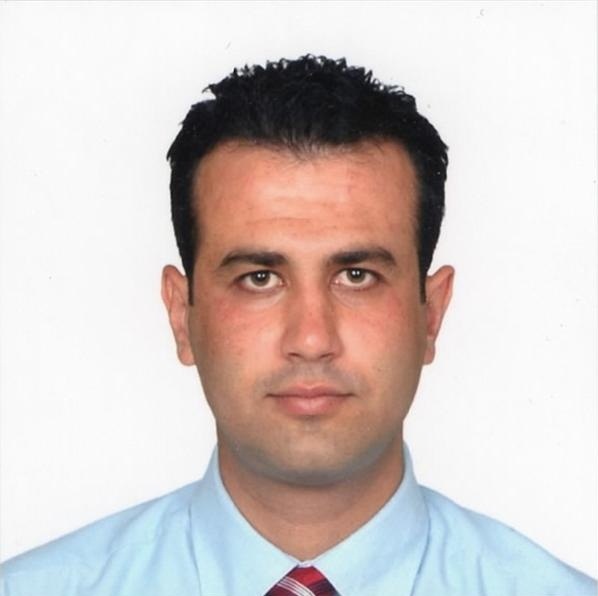}}]{Ali R{\i}za Ekti} is from Tarsus, Turkey. He received the B.Sc. degree in electrical and electronics engineering from Mersin University, Mersin, Turkey, in June 2006, the M.Sc. degree in electrical engineering from the University of South Florida, Tampa, FL, USA, in December 2009, and the Ph.D. degree in electrical engineering from the Department of Electrical Engineering and Computer Science, Texas A\&M University, in August 2015. He was a Visiting Professor at the Electrical and Computer Engineering Department of Gannon University, Erie, PA between August 2015 and June 2016. He worked as an Assistant Professor with the Electrical and Electronics Engineering Department, Balıkesir University and held the Division Manager position of HISAR Laboratory, TÜBİTAK BİLGEM, where he was responsible for research and development activities in the wireless communications and signal processing between November 2016 and December 2021. In January 2022, he joined Grid Communications and Security Research Group, Oak Ridge National Laboratory, Knoxville, TN, USA, where he is currently a Senior R\&D Staff Member. His current research interests include statistical signal processing, wireless propagation channel modeling, optimization, machine learning in 5G and beyond systems and anomaly detection in smart grid.
\end{IEEEbiography}

\begin{IEEEbiography}
    [{\includegraphics[width=1in,height=1.25in,clip,keepaspectratio]{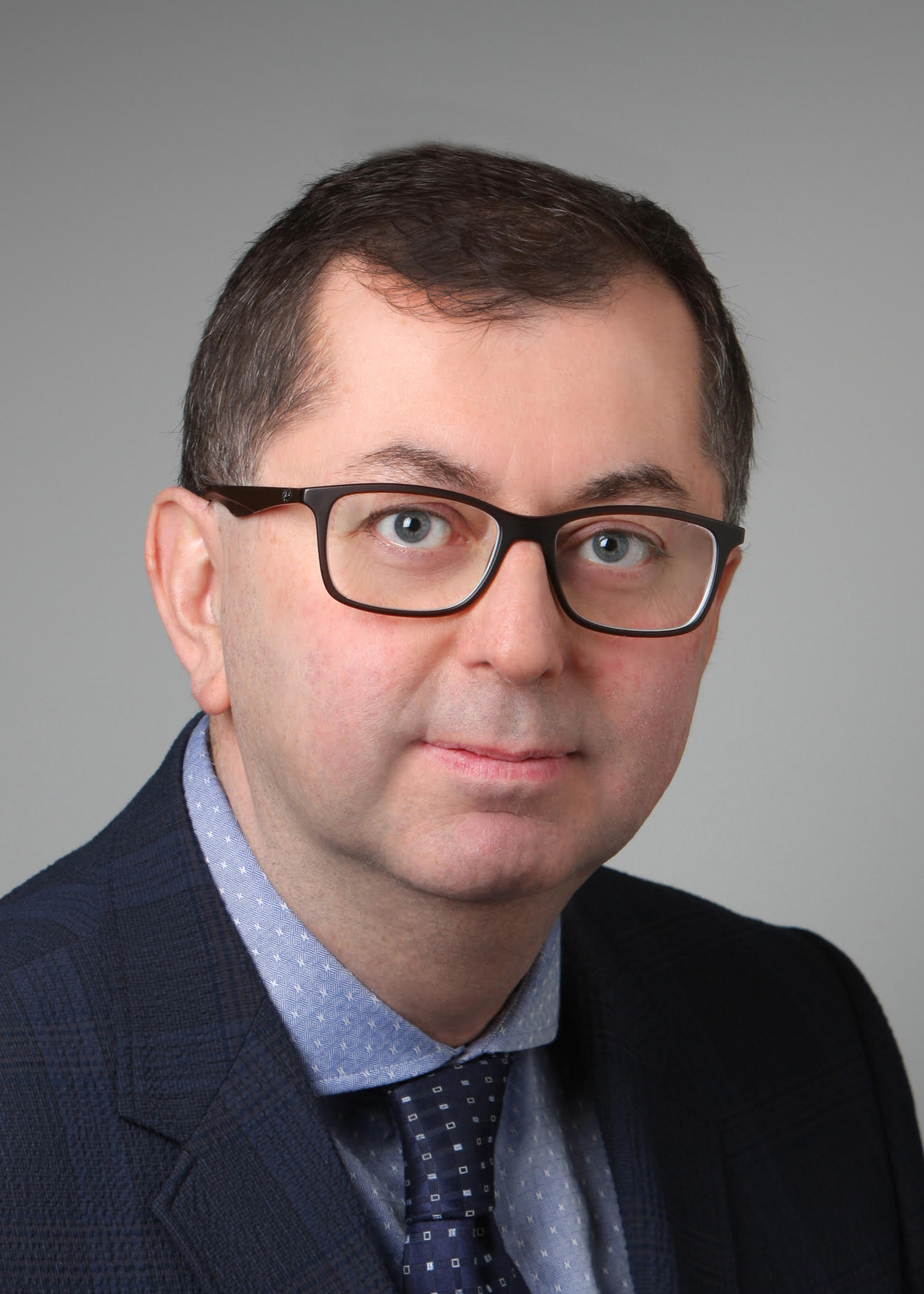}}]{Halim Yanikomeroglu} [F] (halim@sce.carleton.ca) is a full professor in the Department of Systems and Computer Engineering at Carleton University, Ottawa, Canada. His research interests cover many aspects of 5G/5G+ wireless networks. His collaborative research with industry has resulted in 39 granted patents. He is a Fellow of the Engineering Institute of Canada and the Canadian Academy of Engineering, and he is a Distinguished Speaker for IEEE Communications Society and IEEE Vehicular Technology Society.
\end{IEEEbiography}

\end{document}